\documentclass[useAMS,usenatbib]{mn2e}

\usepackage{graphicx}	% Including figure files
\usepackage{amsmath}	% Advanced maths commands
\usepackage{amssymb}	% Extra maths symbols
\usepackage{multicol}   % Multi-column entries in tables
\usepackage{bm}		% Bold maths symbols, including upright Greek
\usepackage{pdflscape}	% Landscape pages
\usepackage{multirow,multicol} 

\def\cm{{\rm\thinspace cm}}

\def\erg{{\rm\thinspace erg}}

\def\Lsun{\hbox{$\rm\thinspace L_{\odot}$}}

\def\Msun{\hbox{$\rm\thinspace M_{\odot}$}}
\def\Zsun{\hbox{$\rm\thinspace Z_{\odot}$}}

\def\s{{\rm\thinspace s}}
\def\yr{{\rm\thinspace yr}}

\def\ergpcmcups{\hbox{$\erg\cm^{-3}\s^{-1}\,$}}
\def\ergpcmcups2{\hbox{$\erg\cm^{3}\s^{-1}\,$}}

\def\Msunpyr{\hbox{$\Msun\yr^{-1}\,$}}

\def\N1316{NGC\,1316}
\def\N1404{NGC\,1404}

\def\4U{4U~1735$-$444}

\def\arcsec{\ifmmode '' \else $''$\fi}

\def\arcsecpoint{\ifmmode ''\!. \else $''\!.$\fi}

\def\kms{\ifmmode {\rm km\ s}^{-1} \else km s$^{-1}$\fi}
\def\Msun{\ifmmode {\rm M}_{\odot} \else M$_{\odot}$\fi}
\def\Lsun{\ifmmode {\rm L}_{\odot} \else L$_{\odot}$\fi}
\def\Zsun{\ifmmode {\rm Z}_{\odot} \else Z$_{\odot}$\fi}

\def\ergscm2{ergs\,s$^{-1}$\,cm$^{-2}$}
\def\icm3{{\rm cm}^{-3}}
\def\icm2{{\rm cm}^{-2}}
\def\qo{\ifmmode q_{\rm o} \else $q_{\rm o}$\fi}
\def\Ho{\ifmmode H_{\rm o} \else $H_{\rm o}$\fi}
\def\ho{\ifmmode h_{\rm o} \else $h_{\rm o}$\fi}

\def\vFWHM{\ifmmode v_{\mbox{\tiny FWHM}} \else
            $v_{\mbox{\tiny FWHM}}$\fi}
\def\CCF{\ifmmode F_{\it CCF} \else $F_{\it CCF}$\fi}
\def\ACF{\ifmmode F_{\it ACF} \else $F_{\it ACF}$\fi}
\def\Halpha{\ifmmode {\rm H}\alpha \else H$\alpha$\fi}
\def\Hbeta{\ifmmode {\rm H}\beta \else H$\beta$\fi}
\def\Hgamma{\ifmmode {\rm H}\gamma \else H$\gamma$\fi}
\def\Hdelta{\ifmmode {\rm H}\delta \else H$\delta$\fi}
\def\Lya{\ifmmode {\rm Ly}\alpha \else Ly$\alpha$\fi}
\def\Lyb{\ifmmode {\rm Ly}\beta \else Ly$\beta$\fi}
\def\Lyg{\ifmmode {\rm Ly}\beta \else Ly$\gamma$\fi}

\def\ciii{\ifmmode {\rm C}\,{\sc iii} \else C\,{\sc iii}\fi}
\def\civ{\ifmmode {\rm C}\,{\sc iv} \else C\,{\sc iv}\fi}
\def\cv{\ifmmode {\rm C}\,{\sc v} \else C\,{\sc v}\fi}
\def\cvi{\ifmmode {\rm C}\,{\sc vi} \else C\,{\sc vi}\fi}

\def\o5007{[O\,{\sc iii}]\,$\lambda5007$}

\def\fexxii-iii{Fe\,{\sc xxii-xxiii}}

\usepackage{hyperref}
\hypersetup{colorlinks=true,linkcolor=blue,citecolor=blue,filecolor=blue,urlcolor=blue}

\title[Cooling gas in galaxy clusters and groups]
  {Searching for cool and cooling X-ray emitting gas in 45 galaxy clusters and groups} 
  \author[Haonan Liu et al.]{Haonan Liu,$^{1}$ \thanks{E-mail: hl479@cam.ac.uk} 
  Ciro Pinto,$^{1,2}$ Andrew C. Fabian,$^{1}$ Helen R. Russell $^{1}$
  \newauthor and Jeremy S. Sanders\,$^{3}$ \\
  $^{1}$Institute of Astronomy, Madingley Road, CB3 0HA Cambridge, United Kingdom\\
  $^{2}$ESTEC/ESA, Keplerlaan 1, 2201AZ Noordwijk, The Netherlands\\
  $^{3}$Max-Planck-Institut f{\"u}r extraterrestrische Physik, Giessenbachstrasse, 85748 Garching, Germany\\
}
\begin{document}

\date{\today}

\pagerange{\pageref{firstpage}--\pageref{lastpage}} \pubyear{2018}

\maketitle

\label{firstpage}

\begin{abstract}
  We present a spectral analysis of cool and cooling gas in 45 cool-core clusters and groups of galaxies obtained from Reflection Grating Spectrometer (RGS) XMM-\textit{Newton} observations. 
  The high-resolution spectra show {Fe\,\scriptsize{XVII}} emission in many clusters, which implies the existence of cooling flows. 
  The cooling rates are measured between the bulk Intracluster Medium (ICM) temperature and 0.01 keV and are typically weak, 
  operating at less than a few tens of $\Msunpyr$ in clusters, and less than 1 $\Msunpyr$ in groups of galaxies. 
  They are 10-30$\%$ of the classical cooling rates in the absence of heating, which suggests that AGN feedback has a high level of efficiency. 
  If cooling flows terminate at 0.7 keV in clusters, the associated cooling rates are higher, and have a typical value of a few to a few tens of $\Msunpyr$. 
  Since the soft X-ray emitting region, where the temperature $kT<1$ keV, is spatially associated with H$\alpha$ nebulosity, we examine the relation between the cooling rates above 0.7 keV and the H$\alpha$ nebulae. 
  We find that the cooling rates have enough energy to power the total UV-optical luminosities, and are 5 to 50 times higher than the observed star formation rates for low luminosity objects. 
  In 4 high luminosity clusters, the cooling rates above 0.7 keV are not sufficient and an inflow at a higher temperature is required. 
  Further residual cooling below 0.7 keV indicates very low complete cooling rates in most clusters.

\end{abstract}

\begin{keywords}
X-rays: galaxies: clusters - galaxies: clusters: general
\end{keywords}

\section{Introduction}
\label{sec:intro}

The centre of gravitational systems is one of the key aspects in understanding structure formation.  
In a hierarchical formation scheme, more massive structures form through merging of smaller components in overdense regions.  
This self-similar behaviour implies that the dominance of dark matter potential wells in galaxy clusters leads to an inflow of baryons which will deposit the gravitational energy in the core.  
In hydrostatic equilibrium, such a system will have a high gas temperature and pressure, while preventing overdensity in the central region.  
However, it is realised that the evolution of galaxy clusters involves processes other than gravitational collapse, 
such as cooling and feedback (e.g. \citealt{1991ApJ...383..104K};  \citealt{2000MNRAS.318..889W}; \citealt{2002ApJ...576..601V}).  
Most evidently, a large fraction of galaxy clusters have been found to host cool cores where the temperature drops towards the centre 
(e.g. \citealt{1984ApJ...285....1S}; \citealt{2005MNRAS.359.1481B}; \citealt{2009ApJS..182...12C}; \citealt{2010A&A...513A..37H}).
The central radiative cooling time drops below a few $10^8$ yr (e.g. \citealt{2002MNRAS.332L..50F}), 
and the entropy $K$ also decreases inwards by a power law (\citealt{2008ApJ...683L.107C,2009ApJS..182...12C}; \citealt{2014MNRAS.438.2341P}).
These suggest that a radiative cooling flow forms in the central region (\citealt{1994ARA&A..32..277F}), 
where the energy loss can be observed directly in X-rays by thermal bremsstrahlung.
In the cooling flow model, cool gas is compressed by the weight of overlaying gas,
and a subsonic inflow of hot gas from outer region is required to sustain pressure.
In the absence of heating, cooling rates are predicted to be 100s to more than 1000 $\rm M_{\odot}\,\rm yr^{-1}$ in rich clusters 
(\citealt{1997MNRAS.292..419W}; \citealt{1998MNRAS.298..416P}; \citealt{2001MNRAS.322..589A}; \citealt{2010A&A...513A..37H}; \citealt{2018ApJ...858...45M}).
This suggests that we expect not only low temperature components in X-rays but also a large amount of cold molecular gas if it is not consumed in star formation.

On the contrary, observations have shown that the star formation rate is only a small fraction of the predicted cooling rate 
(\citealt{1987PASAu...7..132N}; \citealt{1987MNRAS.224...75J}; \citealt{2008ApJ...681.1035O}; \citealt{2008ApJ...687..899R}; \citealt{2018ApJ...858...45M}),
and the molecular gas detected by CO line emission (\citealt{2001MNRAS.328..762E}; \citealt{2003A&A...412..657S}) is at least 20 times lower. 
Meanwhile, far less cooling gas is observed below 1-2 keV in rich clusters (e.g. \citealt{2001A&A...365L..99K}; \citealt{2001A&A...365L.104P}; \citealt{2001A&A...365L..87T}; \citealt{2001ApJ...557..546D}). 
\citealt{2003ApJ...590..207P} demonstrated that the standard cooling flow model overpredicts the emission lines from the lowest temperatures in X-rays. 
The analysis of the Centaurus cluster showed that the cooling rate below 0.8 keV is much lower than the cooling rates measured at hotter temperatures (\citealt{2008MNRAS.385.1186S})
; a similar result was obtained for M87 by \citet{2010MNRAS.407.2063W}.
Therefore, cooling must be suppressed by heating mechanisms. 
AGN feedback is the most likely mechanism, which is energetically strong enough to prevent cooling and yet not overheat the core 
(for reviews in AGN feedback, see \citealt{2007ARA&A..45..117M,2012NJPh...14e5023M} and \citealt{2012ARA&A..50..455F}). 
The energy transport mechanism is still uncertain, which should distribute heat spatially within a few tens of kpc. 
Some possible processes are sound waves and gravity waves (see e.g. \citealt{2005MNRAS.363..891F,2017MNRAS.464L...1F}).
Other mechanisms such as dissipation through turbulence and conduction are found to be insufficient to operate the heating process by themselves
(\citealt{2015A&A...575A..38P,2018MNRAS.480.4113P}; \citealt{2018MNRAS.478L..44B}; \citealt{2004MNRAS.347.1130V}).

On the other hand, we can still detect mild cooling flows at around 0.4-0.8 keV from the {Fe\,\scriptsize{XVII}} line emission seen in some objects (e.g. \citealt{2008MNRAS.385.1186S}).
This suggests that AGN feedback cannot perfectly quench radiative cooling.
Further cooling in X-rays is usually not detected from the {O\,\scriptsize{VII}} emission peaking at around 0.1-0.2 keV, 
though there is evidence of detecting weak {O\,\scriptsize{VII}} emission in less massive clusters and groups of galaxies (\citealt{2011MNRAS.412L..35S}; \citealt{2014A&A...572L...8P,2016MNRAS.461.2077P}). 
This raises the question about whether cooling flows can cool further (\citealt{2002MNRAS.332L..50F}).
From spatially-resolved \textit{Chandra} spectra, soft X-ray emitting regions at these temperatures spatially coincide with cooler ultraviolet/optical line-emitting filaments in massive clusters 
(e.g. \citealt{2001MNRAS.321L..33F}; \citealt{2003MNRAS.344L..48F}; \citealt{2005MNRAS.363..216C}). 
These filaments are highly luminous and most of them have luminosities comparable to their soft X-ray emission. 
This suggests that the cool X-ray emitting gas is likely mixing with cold atomic and molecular line-emitting material.
The thermal energy of the hotter X-ray emitting gas is then rapidly radiated at longer wavelengths, e.g., in UV and optical bands (\citealt{2002MNRAS.332L..50F}). 
To relate properties of optical line-emitting filaments to soft X-ray gas, we convert luminosities into mass cooling rates,

\begin{equation}
\label{equ:1}
\dot M= \frac{2}{3} \times \frac{L\mu m_{\rm p}}{kT}
\end{equation}

\noindent where $\mu$ is the mean particle weight, $m_{\rm p}$ is the proton mass and $k$ is the Boltzmann's constant. 
We ignore the $P$d$V$ work done on the cooling gas. 

In this paper, we primarily focus on measuring the cooling rates of galaxy clusters, and deduce the efficiency of AGN feedback on suppressing cooling. 
It is also interesting to compare the measured cooling rates to the energy required to power the observed luminosities at longer wavelengths in filaments. 
We then search for residual cooling rates below 0.7 keV which determine whether the gas can continue to cool radiatively in X-rays. 
Finally, the cooling rates are then linked to star formation rates, and we wish to know how they contribute to the massive molecular gas reservoir seen. 

Throughout this work, we assume the following cosmological parameters: $H_{0} = 73 \rm \ km^{-1}Mpc^{-1}$, $\Omega_{\rm M} = 0.27$, $\Omega_{\rm \Lambda} = 0.73$. 
The results from literature are corrected using the same cosmology. 
This paper is organized as follows. 
Section \ref{sec:data} provides the observations used in our sample and the data reduction procedure. 
The spectral analysis is presented in section \ref{sec:spectral_analysis},
and we discuss the significance of our result in section \ref{sec:discussion}.
Finally, we present our conclusions in section \ref{sec:conclusion}. 

\section[]{Data}
\label{sec:data}

In this paper, we present our analysis of the soft X-ray spectra of 45 nearby cool-core galaxy clusters and groups, 
including the CHEmical Enrichment RGS Sample (CHEERS) sample and the more distant cluster A1835 (\citealt{2015A&A...575A..38P}; \citealt{2017A&A...607A..98D}). 
The CHEERS project includes clusters, groups and elliptical galaxies with the {O\,\scriptsize{VIII}} line detected at 5 $\sigma$ in the Reflection Grating Spectrometer (RGS) spectra (\citealt{2015A&A...575A..38P}), 
and provides a moderately large sample of objects with deep exposure times. 
Some of the original aims of the CHEERS project were to accurately measure the abundances of key elements, e.g, O and Fe (\citealt{2017A&A...607A..98D}) and constrain the level of turbulence (\citealt{2015A&A...575A..38P}). 
These suggest that the sample is also suitable for measuring the cooling structure of clusters below 1-2 keV, 
since the relevant O and Fe ionisation stages in the soft X-ray band are strong and usually peak at different temperatures. 
Furthermore, the CHEERS sample is relatively complete, which contains all suitable targets with different size and bulk temperature within a low redshift of $z\leq 0.1$.

The observations were made by the XMM-\textit{Newton} satellite, and are listed in Table\,\ref{table:log}. 
The satellite has two different types of X-ray instruments: the Reflection Grating Spectrometer and the European Photon Imaging Camera (EPIC). 
There are two RGS detectors 1 and 2, which are slitless with high spectral solution between 7 and 38 \AA\, (1.77 to 0.33 keV), and we use the spectra from both detectors for spectral analysis. 
The MOS 1 and 2 cameras from EPIC are aligned with their associated RGS detectors and have higher spatial resolution, which are used for imaging. 

\begin{table*}
\caption{XMM-\textit{Newton}/RGS observations and target properties.}  
\vspace{-0.25cm}

\label{table:log}
\renewcommand{\arraystretch}{1.1}
 \begin{center}
%  \small\addtolength{\tabcolsep}{+2pt}
 
\scalebox{0.95}{%
\hspace*{-1cm}\begin{tabular}{c c c c c c c }     
\hline\hline            
Source\,$^{\it a}$            &  Observation ID                           & Total clean time (ks)\,$^{\it b}$& $\it kT_{\rm 1 cie}\,^{\it c}$ & $\it z(D_{L})\,^{\it d}$      & $N_{\rm H,tot}(N_{\rm HI})\,^{\it e}$  \\ \hline                                                                                                                                                                                                   
2A0335+096                    &  0109870101/0201 0147800201               & 120.5                            &  1.49$\pm0.02$                 & 0.0363 (151)                  & 30.7 (17.6)                            \\   
A85                           &  0723802101/22011                         & 195.8                            &  3.10$\pm0.12$                 & 0.0551 (236)                  & 3.10 (2.78)                            \\   
A133                          &  0144310101 0723801301/2001               & 168.1                            &  2.20$\pm0.06$                 & 0.0566 (243)                  & 1.67 (1.59)                            \\   
A262                          &  0109980101/0601 0504780101/0201          & 172.6                            &  1.32$\pm0.01$                 & 0.0174 (72.5)                 & 7.15 (5.67)                            \\   
Perseus 90$\%$ PSF (A426)     &  0085110101/0201 0305780101               & 162.8                            &  1.98$\pm0.03$                 & 0.0179 (74.6)                 & 20.7 (13.6)                            \\   
Perseus 99$\%$ PSF            &                                           &                                  &  1.86$\pm0.02$                 &                               &                                        \\   
A496                          &  0135120201/0801 0506260301/0401          & 141.2                            &  2.11$\pm0.05$                 & 0.0329 (139)                  & 6.12 (3.81)                            \\   
A1795                         &  0097820101                               & 37.8                             &  3.09$\pm0.15$                 & 0.0625 (269)                  & 1.24 (1.19)                            \\   
A1835                         &  0098010101 0147330201 0551830101/0201    & 294.7                            &  3.89$\pm0.27$                 & 0.2532 (1230)                 & 2.24 (2.04)                            \\   
A1991                         &  0145020101                               & 41.6                             &  1.55$\pm0.05$                 & 0.0587 (252)                  & 2.72 (2.46)                            \\   
A2029                         &  0111270201 0551780201/0301/0401/0501     & 155.0                            &  3.45$\pm0.13$                 & 0.0773 (336)                  & 3.70 (3.25)                            \\   
A2052                         &  0109920101 0401520301/0501/0601/080      & 104.3                            &  1.74$\pm0.04$                 & 0.0355 (150)                  & 3.03 (2.71)                            \\                      
                              &  0401520901/1101/1201/1301/1601/1701      &                                  &                                &                               &                                        \\       
A2199                         &  0008030201/0301/0601 0723801101/1201     & 129.7                            &  2.57$\pm0.09$                 & 0.0302 (126)                  & 0.909 (0.888)                          \\                      
A2597                         &  0108460201 0147330101 0723801601/1701    & 163.9                            &  2.48$\pm0.14$                 & 0.0852 (373)                  & 2.75 (2.48)                            \\                     
A2626                         &  0083150201 0148310101                    & 56.4                             &  3.07$\pm0.46$                 & 0.0553 (236)                  & 4.59 (3.82)                            \\   
A3112                         &  0105660101 0603050101/0201               & 173.2                            &  2.60$\pm0.08$                 & 0.0753 (327)                  & 1.38 (1.33)                            \\   
Centaurus (A3526)             &  0046340101 0406200101                    & 152.8                            &  1.33$\pm0.00$                 & 0.0114 (47.2)                 & 12.2 (8.56)                            \\   
A3581                         &  0205990101 0504780301/0401               & 123.8                            &  1.33$\pm0.02$                 & 0.023  (96.2)                 & 5.32 (4.36)                            \\   
A4038                         &  0204460101 0723800801                    & 82.7                             &  2.31$\pm0.12$                 & 0.0282 (118)                  & 1.62 (1.53)                            \\                                                          
A4059                         &  0109950101/0201 0723800901/1001          & 208.2                            &  2.47$\pm0.11$                 & 0.0487 (208)                  & 1.26 (1.21)                            \\   
AS1101                        &  0147800101 0123900101                    & 131.2                            &  1.95$\pm0.04$                 & 0.0580 (249)                  & 1.17 (1.14)                            \\   
AWM7                          &  0135950301 0605540101                    & 158.7                            &  1.72$\pm0.04$                 & 0.0172 (71.8)                 & 11.9 (8.69)                            \\   
EXO0422-086                   &  0300210401                               & 41.1                             &  2.31$\pm0.18$                 & 0.0397 (168)                  & 12.4 (7.86)                            \\   
Fornax (NGC1399)              &  0012830101 0400620101                    & 123.9                            &  0.98$\pm0.00$                 & 0.0046 (19.0)                 & 1.56 (1.5)                             \\   
Hydra A                       &  0109980301 0504260101                    & 110.4                            &  2.44$\pm0.16$                 & 0.0549 (235)                  & 5.53 (4.68)                            \\   
Virgo (M87)                   &  0114120101 0200920101                    & 129.0                            &  1.32$\pm0.00$                 & 0.0043 (16.7)                 & 2.11 (1.94)                            \\   
MKW3s                         &  0109930101 0723801501                    & 145.6                            &  2.29$\pm0.08$                 & 0.0442 (188)                  & 3.00 (2.68)                            \\   
MKW4                          &  0093060101 0723800601/0701               & 110.3                            &  1.44$\pm0.02$                 & 0.02   (83.4)                 & 1.88 (1.75)                            \\ \hline                                                                                                                                                                          
HCG62                         &  0112270701 0504780501 0504780601         & 164.6                            &  0.84$\pm0.01$                 & 0.0147 (61.2)                 & 3.81 (3.31)                            \\   
NGC5044                       &  0037950101 0584680101                    & 127.1                            &  0.87$\pm0.00$                 & 0.0093 (38.4)                 & 6.24 (4.87)                            \\   
NGC5813                       &  0302460101 0554680201/0301/0401          & 146.8                            &  0.68$\pm0.00$                 & 0.0065 (26.9)                 & 5.19 (4.37)                            \\  
NGC5846                       &  0021540101/0501 0723800101/0201          & 162.8                            &  0.70$\pm0.00$                 & 0.0057 (23.6)                 & 5.12 (4.29)                            \\  
M49                           &  0200130101                               & 81.4                             &  0.89$\pm0.01$                 & 0.0033 (16.0)                 & 1.63 (1.53)                            \\ 
M86                           &  0108260201                               & 63.5                             &  0.79$\pm0.01$                 &-0.0008 (16.4)                 & 2.98 (2.67)                            \\   
M89                           &  0141570101                               & 29.1                             &  0.60$\pm0.02$                 & 0.0011 (16.5)                 & 2.96 (2.62)                            \\   
NGC507                        &  0723800301                               & 94.5                             &  1.07$\pm0.01$                 & 0.0165 (68.5)                 & 6.38 (5.25)                            \\   
NGC533                        &  0109860101                               & 34.7                             &  0.89$\pm0.01$                 & 0.0328 (138)                  & 3.38 (3.08)                            \\   
NGC1316                       &  0302780101 0502070201                    & 165.9                            &  0.68$\pm0.01$                 & 0.0059 (24.2)                 & 2.56 (2.4)                             \\   
NGC1404                       &  0304940101                               & 29.2                             &  0.66$\pm0.01$                 & 0.0065 (26.8)                 & 1.57 (1.51)                            \\   
NGC1550                       &  0152150101 0723800401/0501               & 173.4                            &  1.15$\pm0.01$                 & 0.0129 (51.4)                 & 16.2 (10.2)                            \\   
NGC3411                       &  0146510301                               & 27.1                             &  0.91$\pm0.01$                 & 0.0153 (63.5)                 & 4.55 (3.87)                            \\   
NGC4261                       &  0056340101 0502120101                    & 134.9                            &  0.73$\pm0.01$                 & 0.0074 (30.5)                 & 1.86 (1.75)                            \\   
NGC4325                       &  0108860101                               & 21.5                             &  0.89$\pm0.01$                 & 0.0257 (108)                  & 2.54 (2.32)                            \\   
NGC4374                       &  0673310101                               & 91.5                             &  0.68$\pm0.01$                 & 0.0034 (17.0)                 & 3.38 (2.99)                            \\   
NGC4636                       &  0111190101/0201/0501/0701                & 102.5                            &  0.67$\pm0.01$                 & 0.0031 (16.3)                 & 2.07 (1.9)                             \\   
NGC4649                       &  0021540201 0502160101                    & 129.8                            &  0.84$\pm0.01$                 & 0.0037 (16.9)                 & 2.23 (2.04)                            \\   

\hline                
\end{tabular}}
 \end{center}
$^{(a)}$ The horizontal line between MKW4 and HCG62 differentiates clusters (above) and groups of galaxies (below). 
For the Perseus clusters, we extracted both the 90 and 99$\%$ PSF spectra. 
For A1795, we extracted the 97$\%$ PSF spectrum.
$^{(b)}$ RGS net exposure time. 
$^{(c)}$ The best fit temperatures of the 1 \textit{cie} model in keV. 
$^{(d)}$ The redshifts are taken from the NED database (https://ned.ipac.caltech.edu/). 
The luminosity distances in Mpc shown in brackets are either calculated using \citet{2006PASP..118.1711W} (for $z>0.006$) or taken directly from the NED database (for $z<0.006$).
$^{(e)}$ Total ($N_{\rm H,tot}$) and atomic ($N_{\rm HI}$) hydrogen column densities in $10^{20}\rm\, cm^{-2}$ (see http://www.swift.ac.uk/analysis/nhtot/; \citealt{2005A&A...440..775K}; \citealt{2013MNRAS.431..394W}). \\
      \vspace{0.5cm}
\end{table*}

% \vspace{-0.25cm}

\subsection{Data reduction}
\label{sec:data_reduction}
We follow the data reduction procedure used by \citet{2015A&A...575A..38P} with the XMM-\textit{Newton} Science Analysis System (SAS) v 13.5.0. 
The RGS spectra are processed by the SAS task \textit{rgsproc}, which produces necessary event files, spectra and response matrices. 
We use \textit{emproc} for the MOS 1 data, and the SAS task \textit{evselect} extracts light curves from MOS 1 in the 10-12 keV energy band, 
allowing us to correct for the contamination from soft-proton flares.
The light curves are binned in 100 s intervals and all time bins outside the 2\,$\sigma$ level were rejected. 
We also use template background files based on count rates in CCD 9 to create background spectra, where emission from sources is not expected. 

\subsection{RGS spectra and MOS 1 images}
\label{sec:rgs_spectra}

We use the task \textit{rgsproc} while setting the \textit{xpsfincl} mask to include 90$\%$ of the point spread function (PSF) of the first order spectra, 
which corresponds to a narrow 0.8$^{\prime}$ region containing the central core. 
The product spectra are subsequently converted into SPEX usable format through the SPEX task \textit{trafo}. 
In the conversion process, we stacked different exposures with \textit{rgscombine} to produce average spectra with high statistics.

RGS spectra are broadened due to spatial (angular) extent of sources in dispersion direction by $\Delta\lambda = 0.138\Delta\theta/m$\,\AA,
where $\Delta\lambda$ is the wavelength shift, $\Delta\theta$ is the angular offset from the central source in arcmin and $m$ is the spectral order (see the XMM-\textit{Newton} Users Handbook for a complete description).
We expect that spatial broadening is more important for nearby sources, because angular extents depend on redshifts. 
To correct for broadening effects, we need to extract surface brightness profile of all sources from their MOS 1 image in the 0.5-1.8 keV energy band with the task \textit{vprof}, 
where the products are used as the input of the spatial broadening (\textit{lpro}) model in SPEX.
We perform spectral analysis with SPEX version 3.04.00 with its default proto-Solar abundances of \citet{2009M&PSA..72.5154L}. 
In this work, we use C-statistics (C-stat), and adopt $1\,\sigma$ uncertainties ($\Delta$C-stat = 1) for measurements and $2\,\sigma$ uncertainties ($\Delta$C-stat = 2.71) for upper limits, unless otherwise stated. 

\section{Spectral analysis}
\label{sec:spectral_analysis}

In our analysis, we only include the $7-28$\,{\AA} (0.44-1.77 keV) band, since background usually dominates above 28\,{\AA}. 
The spectra are binned by a factor of 5 with the bin size of 0.05\,{\AA}, which ensures that a minimum of $\frac{1}{3}$ of RGS spectral resolution is achieved and the data are not overbinned. 
The typical gas temperature of our low redshift sample is between 0.5 and 4 keV (see Table\,\ref{table:log}). 
In the chosen energy band, we expect emission lines from O, Ne, Mg and Fe with different emissivities. 
For these elements, we set their abundances free relative to hydrogen, except Mg is coupled to Ne to reduce degeneracy in our models which does not significantly change our results. 
The abundance of N cannot be measured precisely, because the background noise is comparable to the {N\,\scriptsize{VII}} emission at rest frame 24.779 {\AA}. 
We choose to couple the abundances of N and all the other elements to Fe. 
These abundances are found to be sub-solar ($Z<Z_{\odot}$) in most objects, though it is possible for the abundance of Fe to be slightly above solar in a few objects such as the Centaurus cluster.  

To search for cool and cooling gas, we model the spectra with collisional ionisation equilibrium (\textit{cie}) and cooling flow (\textit{cf}) components. 
The \textit{cie} component describes an isothermal ICM with a free temperature $T$ and emission measure $EM=n_{\rm e}\,n_{\rm H}\,V$. 
The X-ray luminosities of \textit{cie} components are calculated in the 0.01-10 keV band. 
The \textit{cf} component measures the cooling rate of an isobaric cooling flow from a maximum temperature $T_{\rm max}$ down to a minimum temperature $T_{\rm min}$ (\citealt{Mushotzky}).
Such a cooling rate $\dot{M}$ can be derived from a differential emission measure 

\begin{equation}
\label{equ:0}
\frac{d EM(T)}{d T}= \frac{5\dot{M}k}{2\mu m_{\rm H}\Lambda (T)}
\end{equation}

\noindent where $k$ is the Boltzmann constant, $\mu$ is the mean particle weight, $m_{\rm H}$ is the proton mass and $\Lambda (T)$ is the cooling function. 

Both the \textit{cie} and \textit{cf} components are modified by redshift (\textit{red}), Galactic absorption (\textit{hot}) and then convolved by spatial broadening (\textit{lpro}). 
In the \textit{hot} component, we assume a very cold temperature of $T$=0.5 eV with solar abundances (\citealt{2013A&A...551A..25P}),
and allow hydrogen column density free to vary between $N_{\rm HI}$ and $N_{\rm H,tot}$ (see Table\,\ref{table:log}; \citealt{2005A&A...440..775K}; \citealt{2013MNRAS.431..394W}). 
The hydrogen column densities are crucial in determining the quality of spectral fits and the magnitude of cooling rates (e.g. 2A0335+096). 
Although the intrinsic hydrogen column density of the source can usually be ignored, it can potentially be problematic for clusters with powerful AGN (e.g., the Perseus cluster; \citealt{2003ApJ...590..225C}),
since the emission of the central AGN is processed by the ICM before leaving the clusters. 
These components assume a Maxwellian electron distribution, and calculate the effect of thermal line broadening. 
Although turbulence is intrinsic to ICM, it also broadens emission lines usually at a few 100 km/s (e.g. \citealt{2011MNRAS.410.1797S,2013MNRAS.429.2727S}; \citealt{2015A&A...575A..38P}). 
For the CHEERS sample, the FWHM of total line widths is at least a few 1000 km/s in the RGS spectra (see Table A.1 in \citealt{2015A&A...575A..38P}; \citealt{2016MNRAS.461.2077P}; \citealt{2018MNRAS.478L..44B}). 
Therefore we ignore turbulent velocity in our sample and fit the scaling parameter $s$ in the \textit{lpro} component, which can account for any residual broadening. 
It is possible for our analysis to have minor statistical effects on line widths from stacking multiple observations, though net fluxes are not affected. 
We include an additional power law (\textit{pow}) component for clusters with a bright variable AGN (the Perseus and Virgo clusters; see section \ref{sec:special}). 
The \textit{pow} component is not convolved with the spatial profile as the central AGN is a point source. 
Finally, we assume that any diffuse emission features due to the cosmic X-ray background are smeared out into a broad continuum-like component.

\subsection{Isothermal collisional ionisation equilibrium}
\label{sec:baseline_model}

\begin{figure}
  \includegraphics[width=1.75\linewidth, angle=0]{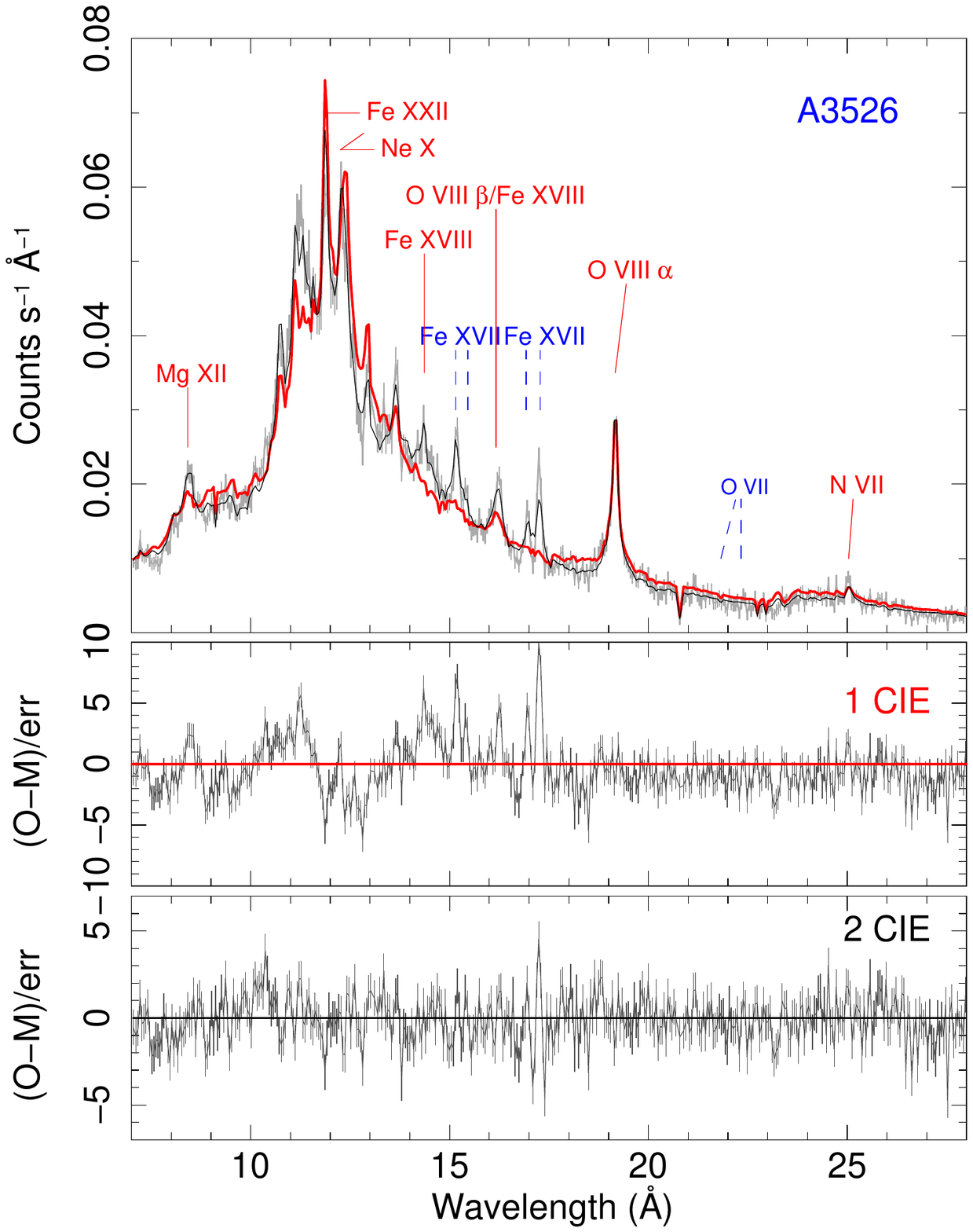}
  \caption{Top: The 90$\%$ PSF RGS spectrum of the Centaurus cluster (grey) with the best-fit isothermal collisional equilibrium model (red) and the 2 temperature model (black).  
  Important emission lines are labelled at the observed wavelengths. 
  Dash lines indicate line emission due to cooling gas below 0.8 keV ({Fe\,\scriptsize{XVII}}) and cooled gas below 0.2 keV ({O\,\scriptsize{VII}}). 
  We show the residuals of both models in the middle (1 \textit{cie}) and in the bottom (2 \textit{cie}) panels.
  Notice that the 2 \textit{cie} model significantly improves the fit to the {Fe\,\scriptsize{XVII}}, {Fe\,\scriptsize{XVIII}} and {Fe\,\scriptsize{XXII}} emission lines.}
  \label{fig:A3526}
\end{figure}

We start with a single collisional ionisation equilibrium component (1 \textit{cie}), and the best fit temperatures are shown in Table\,\ref{table:log}. 
These temperatures are consistently lower than the cluster values listed by \citet{2007A&A...466..805C} and \citet{2008A&A...478..615S}, since the 90$\%$ PSF spectra exclude most of the very hot ($>$4 keV) ICM emission. 
We demonstrate an example spectral fit of the Centaurus cluster in Fig. \ref{fig:A3526}, where we also show the residuals of both the 1 \textit{cie} and 2 \textit{cie} models. 
It is seen that the {Fe\,\scriptsize{XVII/XVIII}} lines between 14 and 17 {\rm \AA} are underestimated in the 1 \textit{cie} model, which gives a poor reduced C-stat of 1952/407. 
The spectral fit is improved significantly the 2 \textit{cie} model, or by adding an additional \textit{cf} component (see \citealt{2008MNRAS.385.1186S} for more detailed analysis on the Centaurus cluster). 
We attempt to trace any cooler component in our sample first by an additional \textit{cie} component at a lower temperature (2 \textit{cie} model; see section \ref{sec:2cie}). 
We then replace this cooler \textit{cie} component with a \textit{cf} component cooling from the hotter \textit{cie} temperature down to 0.01 keV (1 \textit{cie} + 1 \textit{cf} model; section \ref{sec:cooling_flow}). 
Finally, we use a two-stage cooling flow model (1 \textit{cie} + 2 \textit{cf} model; section \ref{sec:cooling_flow}). 

\subsection{Multi-temperature model}
\label{sec:2cie}
The two temperature (2 \textit{cie}) model includes the possibility of a cooler gas component. 
To reduce degeneracy of our model, we assume both \textit{cie} components have the same abundances, 
and are convolved by the same \textit{lpro} component except the Centaurus cluster where the spectral fit is improved by an additional \textit{lpro} component (\citealt{2016MNRAS.461.2077P}). 
This is because the Centaurus cluster has a much smaller extent of cooling gas (\citealt{2008MNRAS.385.1186S,2016MNRAS.457...82S}).  
The 2 \textit{cie} model provides sufficient spectral fits for most clusters which is consistent with \citet{2017A&A...607A..98D}.
In Fig. \ref{fig:A3526}, the example of the Centaurus cluster demonstrates that the cooler \textit{cie} component has a temperature that gives emission from the Fe-L complex and {O\,\scriptsize{VII/VIII}} lines, 
which dominate the total line emissivity below 1 keV (e.g. see Fig. 2 of \citealt{2010MNRAS.402L..11S}). 
The emissivity of each ionisation stage peaks at different temperatures, 
and the best indicators for low temperature gas are {O\,\scriptsize{VII}} lines at around 0.2 keV and {Fe\,\scriptsize{XVII}} lines which have the strongest emissivity below 0.8 keV in the Fe-L complex. 
However, {O\,\scriptsize{VII}} is usually only found in elliptical galaxies but not massive clusters (except e.g. the Centaurus and Perseus clusters; \citealt{2011MNRAS.412L..35S}; \citealt{2016MNRAS.461.2077P}). 
Since the spatial extent of {O\,\scriptsize{VII}} is generally small (e.g. only in the innermost 5 kpc in the Centaurus cluster; \citealt{2016MNRAS.461..922F}), 
it is difficult to measure such lines in our broader 90$\%$ PSF spectra. 
The temperature of the cooler component cannot be constrained from {O\,\scriptsize{VIII}} emission alone, because it has a much wider temperature range. 
Hence, we are mainly interested in detecting cooling gas which emits {Fe\,\scriptsize{XVII}} lines peaking at 0.5 keV.
 
\begin{figure}
  \includegraphics[width=\linewidth, angle=0]{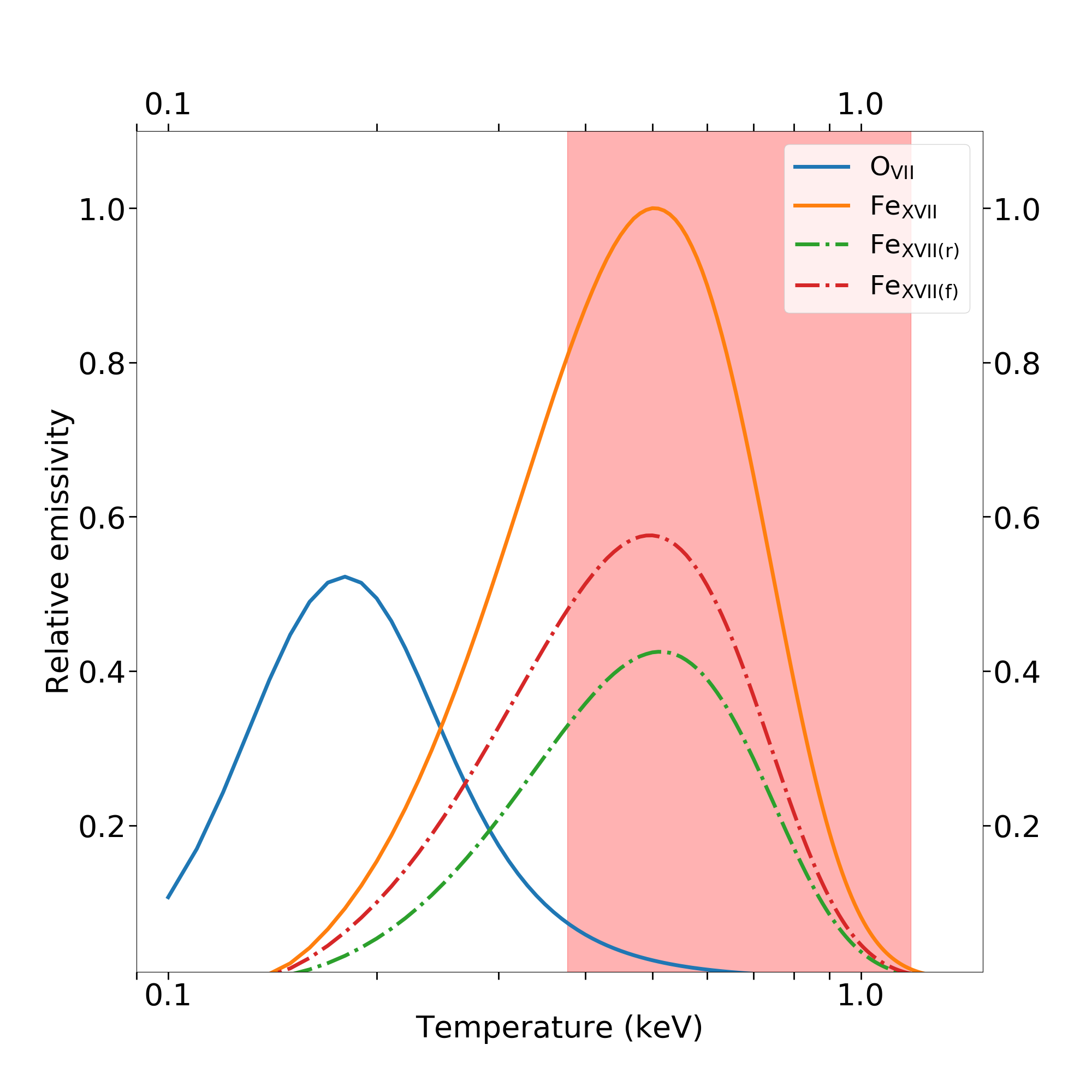}
  \caption{The emissivity of {O\,\scriptsize{VII}} and {Fe\,\scriptsize{XVII}} lines are shown relative to gas temperature. 
   For {Fe\,\scriptsize{XVII}} we include the emissivities of both the resonance and forbidden lines and their sum.  
   The emissivities are normalized such that the peak of the total {Fe\,\scriptsize{XVII}} emissivity is unity. 
   The red shaded region encloses a temperature range 3 times the standard deviation from the average cooler temperature in the 2 \textit{cie} model. 
   This average temperature excludes four objects with their cooler temperature below 0.4 keV.}
  \label{fig:relative}
\end{figure}

We apply the 2 \textit{cie} model to clusters and two bright groups, and the key parameters are listed in Table \ref{table:2cie}. 
There are four objects with a cooler temperature below 0.4 keV, and such a low temperature raises the concern on resonant scattering, 
which can have a significant impact on measuring the gas temperature through certain emission lines. 
Since turbulent velocity is generally low in our sample (\citealt{2015A&A...575A..38P}), the ICM can be optically thick to radiation at resonant lines. 
As a result, photons at the resonant wavelengths are absorbed and re-emitted in random directions, and the resonant lines are suppressed in the core and enhanced from the outer region. 
However, the forbidden line has a much smaller oscillator strength and so is unaffected. 
Consequently, we expect to see a low {Fe\,\scriptsize{XVII}} resonant-to-forbidden ratio in the 90$\%$ PSF spectra. 
We also calculate theoretical emissivities of both the {Fe\,\scriptsize{XVII}} and the {O\,\scriptsize{VII}} lines in Fig. \ref{fig:relative}. 
The resonant-to-forbidden ratio decreases monotonically below around 1 keV with decreasing temperature, and reaches 0.5 at 0.18 keV where the {O\,\scriptsize{VII}} peaks. 
In \citet{2016MNRAS.461.2077P}, it is shown that the {Fe\,\scriptsize{XVII}} resonant-to-forbidden ratio is usually 0.7 or lower. 
Hence, such a low resonance-to-forbidden ratio can be achieved by either a cool ($<$0.2 keV) component or a cooling ($\sim$0.7 keV) component with resonant scattering or a combination of these situations.
Since no {O\,\scriptsize{VII}} lines are observed in most clusters (\citealt{2016MNRAS.461.2077P}), it suggests the cool temperature ($<$0.2 keV) in some objects are likely spurious driven by resonant scattering. 
It is also possible that background subtraction is affecting the spectra.

\begin{figure}
  \includegraphics[width=\linewidth, angle=0]{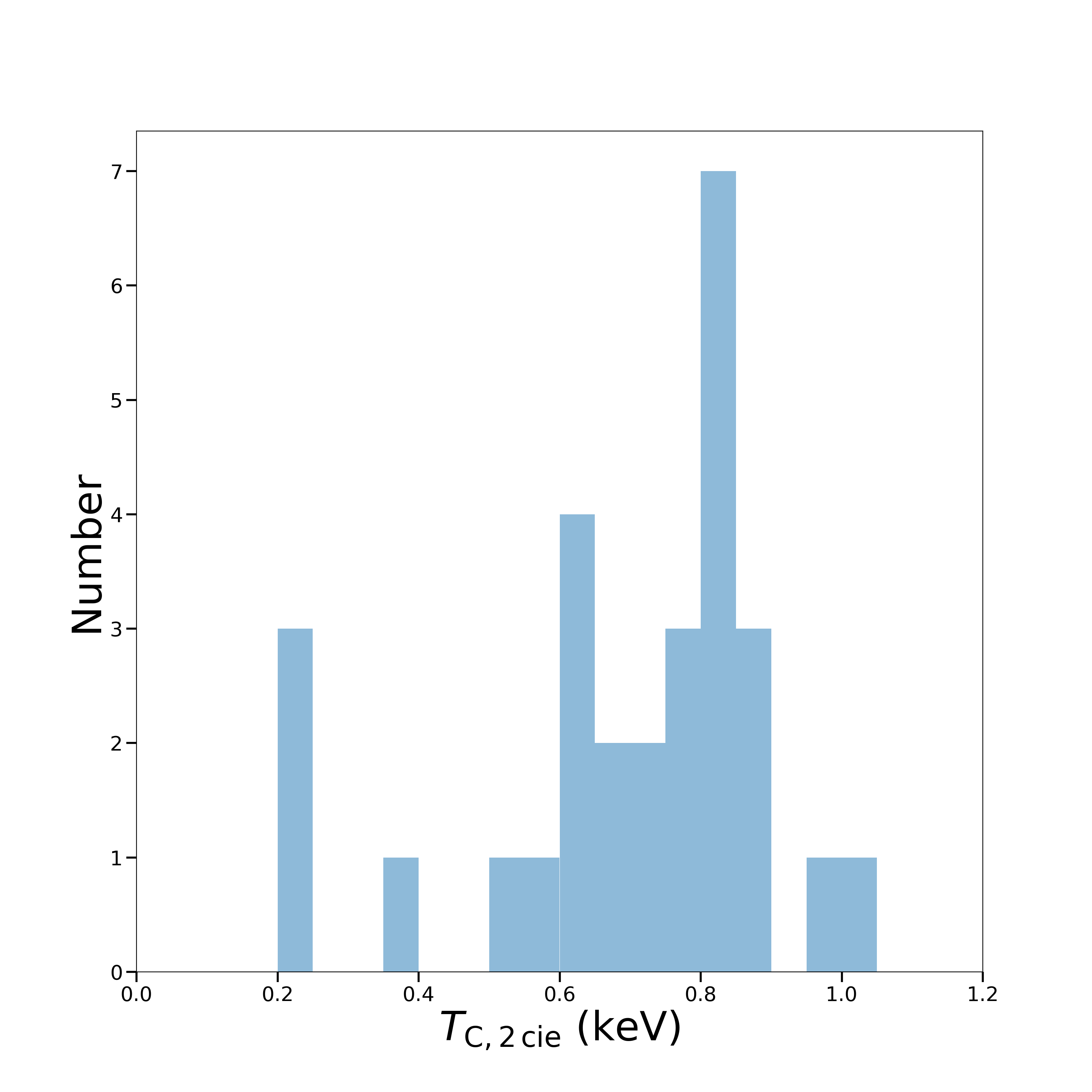}
  \caption{The distribution of the cooler temperature in the 2 \textit{cie} model in rich clusters with a bin size of 0.05 keV.}
  \label{fig:t_hist}
\end{figure}

Excluding the four objects with a cooler temperature below 0.4 keV, the average temperature of the cooler component is determined to be 0.78$\pm$0.13 keV. 
We do not expect the intrinsic temperature of the cooler component to distribute much beyond 3 times the standard deviation from the average, or equivalently below a minimum temperature of 0.39 keV. 
The distribution of the cooler \textit{cie} temperature is seen in Fig. \ref{fig:t_hist}. 
For EXO0422-086, there is a large uncertainty in the cooler temperature due to limited statistics and the luminosity of the same component only gives an upper limit. 
Therefore it does not violate our simple expectation of a minimum cooler temperature of 0.39 keV. 
The best fit models of A1795, AS1101 and MKW3s all give $T_{\rm C, 2\, cie}$ at around 0.22 keV. 
These temperatures are inconsistent with our expectation, and the luminosities also give upper limits. 
We conclude that these measurements are affected by resonant scattering. 
For clusters marked by $*$, no H$\alpha$ filament is detect in A2029 (\citealt{2005MNRAS.360..748J}), hence we strongly suspect it has no cooler \textit{cie} component. 
Both A2626 and Hydra A have limited statistics and the 1 \textit{cie} model can fit their spectra well (\citealt{2017A&A...607A..98D}). 
Although there is a large uncertainty in $T_{\rm H, 2\, cie}$ in Fornax, the 2 \textit{cie} model improves the spectral fit significantly.

The X-ray luminosity of the cooler component is usually $10^{42}-10^{43}\,\rm erg\,s^{-1}$. 
Using equation \ref{equ:0}, these luminosities are equivalent to mass flow rates, which are less than 25 $\Msunpyr$ in most objects. 
This allows us to estimate the volume occupied by the cooler component, the associated gas mass and the cooling time if we know the associated electron density. 
\citet{2009ApJS..182...12C} provided spatially resolved analysis of clusters, where they measured the temperatures and electron densities of gas at hot phase. 
We extrapolate/interpolate these profiles, and evaluate the temperature $T_{\rm e}$ and density $n_{\rm e}$ at a fiducial radius of 5 kpc from the centre. 
This approximation gives an estimated 5$\%$ uncertainties in both quantities. 
By assuming the hotter gas is at pressure equilibrium with the cooler \textit{cie} component, we can estimate the electron density $n_{\rm C}$ associated with the cooler gas by
$n_{\rm e}T_{\rm e}=n_{\rm C}T_{\rm C, 2\, cie}$. 
The volume of the component can then be easily evaluated by emission measure in SPEX, which is the product of the electron and hydrogen densities and the volume. 
If the cool gas is spherical and has an approximately constant density, we can calculate the filling radius $R_{\rm C, 2\, cie}$ and the mass of the cooler gas by $M_{\rm C, 2\, cie}=1.14\times n_{\rm C}m_{\rm H}V$, 
where we assume the hydrogen fraction is 75$\%$ and $V$ is the volume. 
In most objects, we find the filling radius less than 5 kpc and the volume filling ratio of 10-20$\%$. 
It implies that if the cool gas were distributing throughout the 5 kpc core, the gas has to form either narrow filaments or several separated gas clouds. 
There are a few objects with the volume filling ratio larger than 100$\%$, which suggests a larger fiducial radius. 
Since electron densities decrease with larger radii, it is uncertain whether the filling ratio can drop below unity. 
It is likely that the cooler component in these objects are very extended. 
On the other hand, we deduce that the mass of the cooler gas is of the order of $10^{8}-10^{9}\,\Msunpyr$. 
For A262 and 2A0335+096, we find this mass consistent with the molecular mass within a factor of 2 (\citealt{2001MNRAS.328..762E}; Russell et al. in prep). 
However, the molecular mass is not consistent with $M_{\rm C, 2\, cie}$ for other objects. 
Finally, we define the cooling time of the cooler \textit{cie} component to be $t_{\rm cool, 2\, cie}=2.31\times3kT_{\rm C, 2\, cie}/2n_{\rm C}\Lambda$, 
where we get the factor 2.31 by assuming the total hydrogen and ion density is 0.92$n_{\rm C}$, and the proton density is 0.83 $n_{\rm C}$ (\citealt{2018ApJ...858...45M}).
This is larger but proportional to $M_{\rm C, 2\, cie}/\dot M_{\rm C, 2\, cie}$ using our assumptions.

We do not use the same 2 \textit{cie} model on most groups of galaxies, where some objects can be well described by an isothermal ICM (e.g. NGC3411). 
From the improvement of reduced C-stat, we find that a few more groups can be fitted by the 2 \textit{cie} model, which is mostly consistent with \citet{2017A&A...607A..98D}.
Since these objects typically have 1 \textit{cie} temperatures less than 1 keV, it is difficult to resolve the temperature of the additional component. 
Additionally, the cooler \textit{cie} may suppress the original component and force it to have an unexpectedly high temperature in some objects. 

We simulate a spectrum of a cooling flow model from 2 down to 0.01 keV, which is then fitted by a \textit{cie} component. 
This \textit{cie} temperature is found to be 0.86 keV, in agreement with the average temperature of the cooler component. 
\citet{2010MNRAS.402..127S} performed a Markov Chain Monte Carlo analysis on a simulated cooling flow with three variable temperature components and one component at a fixed temperature. 
The distribution also showed that there is a component at 0.6-0.8 keV. 
\citet{2010MNRAS.402..127S} suggested that this particular temperature range is due to gas temperatures which are easily differentiated spectrally. 
These simulations suggest that the cooler \textit{cie} component can instead be a cooling flow in clusters. 
We attempt to trace such a cooling flow with two different models in Section \ref{sec:cooling_flow}.

\begin{table*}
\caption{Key parameters of the 2 \textit{cie} model in clusters and two bright groups.}  
\vspace{-0.25cm}
\label{table:2cie}
\renewcommand{\arraystretch}{1.1}
 \begin{center}
 \small\addtolength{\tabcolsep}{+2pt}
 
\scalebox{1}{%
\hspace{-2cm}\begin{tabular}{c c c c c c c c c c c}     
\hline\hline            
Source                 &  $T_{\rm H, 2\, cie}$& $T_{\rm C, 2\, cie}$& $L_{\rm C, 2\, cie}$& $\dot M_{\rm C, 2\, cie}$ & $R_{\rm C, 2\, cie}$ & $M_{\rm C, 2\, cie}$ & $t_{\rm cool, 2\, cie}$\\
\hline                                                                                                    
2A0335+096             &  1.71$\pm 0.04$         & 0.79$\pm 0.03$         & 650$\pm 60$          & 54$\pm 5$                & 5.3$\pm0.3$     & 2.8$\pm0.6$     &  43$\pm6$          \\
A85                    &  3.14$\pm 0.12$         & 0.82$\pm 0.14$         & $<$126               & $<$10.2                  & $<$2.15         & $<$0.331        &  $>$26.9           \\
A133                   &  2.37$\pm 0.10$         & 0.86$\pm 0.06$         & 160$\pm 40$          & 12$\pm 3$                & 4.2$\pm0.5$     & 0.8$\pm0.4$     &  60$\pm20$         \\
A262                   &  1.52$\pm 0.03$         & 0.79$\pm 0.03$         & 49$\pm 6$            & 4.1$\pm 0.5$             & 4.8$\pm0.4$     & 0.7$\pm0.2$     &  130$\pm20$        \\
Perseus 90$\%$ PSF     &  2.25$\pm 0.05$         & 0.57$\pm 0.03$         & 190$\pm 10$          & 22$\pm 2$                & 2.2$\pm0.1$     & 0.50$\pm0.1$    &  18$\pm2$          \\
Perseus 99$\%$ PSF     &  2.07$\pm 0.03$         & 0.62$\pm 0.02$         & 600$\pm 20$          & 64$\pm 3$                & 3.3$\pm0.2$     & 1.6$\pm0.3$     &  20$\pm2$          \\
A496                   &  2.21$\pm 0.07$         & 0.83$\pm 0.06$         & 84$\pm 10$           & 6.7$\pm 0.8$             & 2.7$\pm0.2$     & 0.40$\pm0.1$    &  47$\pm8$          \\
A1795                  &  3.22$\pm 0.18$         & 0.18$\pm 0.03$         & 800$\pm 300$         & 300$\pm 100$             & 1.8$\pm0.4$     & 0.6$\pm0.4$     &  2$\pm1$           \\
A1835                  &  4.14$\pm 0.18$         & 0.73$\pm 0.07$         & 1600$\pm 500$        & 150$\pm 40$              & 2.8$\pm0.4$     & 1.9$\pm0.9$     &  11$\pm5$          \\
A1991                  &  1.78$\pm 0.11$         & 0.83$\pm 0.07$         & 270$\pm 80$          & 21$\pm 6$                & 5.8$\pm0.9$     & 2.0$\pm0.9$     &  80$\pm30$         \\
A2029 *                &  3.46$\pm 0.16$         & 0.64$\pm_{0.63}^{0.86}$& $<$575               & $<$59.6                  & $<$2.19         & $<$0.908        &  $>$12.6           \\
A2052                  &  1.98$\pm 0.07$         & 0.89$\pm 0.04$         & 180$\pm 40$          & 13$\pm 3$                & 6.0$\pm0.6$     & 1.8$\pm0.6$     &  110$\pm30$        \\
A2199                  &  2.66$\pm 0.10$         & 0.74$\pm 0.06$         & 60$\pm 10$           & 5$\pm 1$                 & 2.3$\pm0.3$     & 0.25$\pm0.10$   &  40$\pm10$         \\
A2597                  &  2.61$\pm 0.10$         & 0.83$\pm 0.09$         & 300$\pm 100$         & 22$\pm 9$                & 3.4$\pm0.7$     & 1.0$\pm0.7$     &  40$\pm20$         \\
A2626 *                &  3.28$\pm_{0.43}^{0.58}$& 1.02$\pm_{0.33}^{0.48}$& $<$83.8              & $<$5.47                  & $<$8.55         & $<$5.16         &  $>$111            \\                                                                                                                                                                          
A3112                  &  2.64$\pm 0.09$         &0.65$\pm 0.08$          & 100$\pm 40$          & 10$\pm 4$                & 2.0$\pm0.4$     & 0.26$\pm0.16$   &  20$\pm10$         \\
Centaurus              &  1.64$\pm 0.02$         &0.82$\pm 0.01$          & 77$\pm 2$            & 6.2$\pm 0.2$             & 3.2$\pm0.2$     & 0.38$\pm0.06$   &  50$\pm3$          \\
A3581                  &  1.38$\pm 0.02$         &0.62$\pm 0.07$          & 37$\pm 7$            & 3.9$\pm 0.7$             & 3.4$\pm0.4$     & 0.4$\pm0.1$     &  80$\pm20$         \\
A4038                  &  2.40$\pm 0.14$         &0.54$\pm 0.10$          & 25$\pm 7$            & 3.1$\pm 0.8$             & 2.4$\pm0.4$     & 0.2$\pm0.1$     &  50$\pm20$         \\
A4059                  &  2.55$\pm 0.12$         &0.84$\pm 0.09$          & 60$\pm 20$           & 4$\pm 2$                 & 3.5$\pm0.7$     & 0.4$\pm0.3$     &  80$\pm50$         \\
AS1101                 &  1.96$\pm 0.05$         &0.23$\pm 0.15$          & $<$129               & $<$36.6                  & $<$1.61         & $<$0.244        &  $>$5.51           \\
AWM7                   &  1.98$\pm 0.10$         &0.68$\pm 0.04$          & 28$\pm 4$            & 2.8$\pm 0.4$             & 2.7$\pm0.2$     & 0.23$\pm0.06$   &  70$\pm10$         \\
EXO0422-086            &  2.30$\pm 0.16$         &0.37$\pm_{0.14}^{0.27}$ & $<$59.4              & $<$10.6                  & $<$1.61         & $<$0.106        &  $>$8.25           \\
Fornax                 &  2.70$\pm_{0.63}^{0.77}$&0.95$\pm 0.01$          & 13$\pm 2$            & 0.9$\pm 0.1$             & /               & /               &  /                 \\
Hydra A *              &  2.44$\pm 0.09$         &0.62$\pm_{0.61}^{0.88}$ & $<$79.8              & $<$8.46                  & $<$2.09         & $<$0.297        &  $>$29.0           \\
Virgo                  &  1.42$\pm 0.01$         &0.80$\pm 0.02$          & 13$\pm 1$            & 1.1$\pm 0.1$             & 2.2$\pm0.2$     & 0.11$\pm0.02$   &  80$\pm10$         \\
MKW3s                  &  2.35$\pm 0.09$         &0.21$\pm 0.05$          & 120$\pm 60$          & 40$\pm 20$               & 1.7$\pm0.5$     & 0.2$\pm0.2$     &  5$\pm4$           \\
MKW4                   &  1.67$\pm 0.16$         &1.11$\pm 0.10$          & 60$\pm 20$           & 3$\pm 1$                 & 5.2$\pm0.9$     & 0.7$\pm0.4$     &  170$\pm90$        \\
HCG62                  &  1.22$\pm 0.10$         &0.78$\pm 0.02$          & 74$\pm 8$            & 6.3$\pm 0.6$             & 9.1$\pm0.6$     & 2.1$\pm0.5$     &  270$\pm40$        \\
NGC5044                &  1.27$\pm_{0.16}^{0.29}$&0.86$\pm 0.01$          & 164$\pm 10$          & 12.7$\pm 0.8$            & 10.8$\pm0.6$    & 4.2$\pm0.8$     &  270$\pm30$        \\

\hline                                                                                                                                                                                                                                           
\end{tabular}}                                                                                                                                                                                                                                   
 \end{center}                                                                                                                                                                                                                                    
We define the condition $\frac{x}{\sigma_{x}}>2$ for a value $x$ (except temperature) to be a measurement, otherwise it is considered as an upper limit. 
This rule does not apply to the last three column due to rounding. 
The best fit temperatures of the 2 \textit{cie} model measured in keV. 
The luminosities of the cooler \textit{cie} component $L_{\rm C, 2\, cie}$ are calculated in the 0.01-10 keV energy band in $10^{40}\,\rm erg\,s^{-1}$, 
and are converted into mass flow rates $\dot M_{\rm C, 2\, cie}$ in $\rm M_{\odot}\,\rm yr^{-1}$ using equation \ref{equ:1} (\citealt{2002MNRAS.332L..50F}).
Assuming the cool gas at $T_{\rm C, 2\, cie}$ is at pressure equilibrium with the hotter gas, it can fill a sphere with effective radius of $R_{\rm C, 2\, cie}$ in kpc. 
Such a sphere with constant density contains the mass of the cool gas $M_{\rm C, 2\, cie}$ in $10^9\,\Msunpyr$. 
The cooling time of the cool gas is also included in $\rm Myr$. 
Clusters and two bright groups are included in this table, and those marked by $*$ usually have high uncertainties in $T_{\rm H, 2\, cie}$ or $T_{\rm C, 2\, cie}$ and $L_{\rm C, 2\, cie}$ gives upper limits.\\                                                                                                
                                                                                                                                                               
\vspace{0.5cm}                                                                                                                                                                                                                                   
\end{table*}                                                                                                                                                                                                                                                                                                   
                                                                                                                                                                                                                                                 
 \vspace{-0.25cm}

\subsection{Cooling flow models}                                                                                                                             
\label{sec:cooling_flow}
\subsubsection{One-stage cooling flow model}
\begin{table*}
\caption{Key parameters of the one-stage and two-stage cooling flow models in clusters and groups.}  
 \vspace{-0.25cm}
\label{table:2cf}     
\renewcommand{\arraystretch}{1.1}

\scalebox{1}{%
\hspace*{-1cm}\begin{tabular}{c c c c c c c c c c}     
\hline\hline            
Source           & $\dot M_{\rm simple, 3Gyr}$ & $\dot M_{\rm simple, 7.7Gyr}$  & $\dot M_{\rm 1 cie + 1 cf}$  & $\dot M_{\rm H, \,1 cie + 2 cf}$  & $\dot M_{\rm C, \,1 cie + 2 cf}$& $L_{\rm H \alpha}$& $\dot M_{\rm neb}$& $\dot M_{\rm SFR}$        & Reference   \\
\hline
2A0335+096            & 112  &185    & 36$\pm 3$                &66$\pm6  $       & 12$\pm4$           & 77.6          & 110                   & 0.4$\pm_{0.1}^{0.2}$      & [1]         \\         
A85                   & 81.7 &142    & 1.3$\pm 0.7$             &$<$4.15          & $<$2.82            & 1.52          & 2.15                  & 0.1$\pm_{0.1}^{2.5}$      & [2]         \\         
A133                  & 47.7 &62.5   & 4.3$\pm 0.5$             &12$\pm2 $        & $<$1.84            & 1.14          & 1.61                  & 0.2$\pm_{0.2}^{1.4}$      & [2]         \\         
A262                  & 3.38 &11.4   & 2.4$\pm 0.2$             &5.3$\pm0.5 $     & 0.9$\pm0.3$        & 0.94          & 1.33                  & 0.21$\pm 0.03$            & [3]         \\         
Perseus 90$\%$ PSF    & 306  &303    & 18$\pm 6$                &$<$5.33          & 31$\pm2$           & 224           & 317                   & 70$\pm_{30}^{60}$         & [4]         \\         
Perseus 99$\%$ PSF    & 306  &303    & 56$\pm 15$               &21$\pm7 $        & 82$\pm4$           & 224           & 317                   & 70$\pm_{30}^{60}$         & [4]         \\         
A496                  & 47.6 &66.9   & 3.2$\pm 0.5$             &8$\pm2 $         & $<$1.92            & 2.95          & 4.17                  & 0.18$\pm 0.01$            & [2]         \\         
A1795                 & 140  &224    & $<$14.9                  &$<$19.5          & $<$21.9            & 5.85          & 8.27                  & 3$\pm_{3}^{13}$           & [3]         \\         
A1835                 & 1010 &1080   & 80$\pm 30$               &$<$114.8         & $<$97.3            & 441           & 624                   & 110$\pm_{40}^{60}$        & [5]         \\         
A1991                 & 37.6 &45.1   & 10$\pm 2$                &24$\pm5$         & $<$4.80            & 3.80          & 5.37                  & 0.7$\pm_{0.5}^{2.8}$      & [2]         \\         
A2029                 & 204  &369    & $<$14.8                  &$<$30.5          & $<$24.3            & $<$4.8        & $<$6.79               & 0.8$\pm_{0.09}^{0.10}$    & [6]         \\         
A2052                 & 28.9 &53.6   & 5.7$\pm 0.6$             &16$\pm1 $        & $<$1.32            & 1.69          & 2.39                  & 0.4$\pm_{0.3}^{1.0}$      & [2]         \\         
A2199                 & 31.8 &107    & 3.2$\pm 0.6$             &5$\pm2 $         & 2$\pm1$            & 1.32          & 1.87                  & 1$\pm_{1}^{9}$            & [3]         \\              
A2597                 & 279  &611    & 11$\pm 4$                &30$\pm10$        & $<$12.3            & 79.1          & 112                   & 4$\pm_{2}^{5}$            & [6]         \\                   
A2626                 & 15.4 &24.8   & 0.9$\pm 0.7$             &3$\pm2$          & $<$2.18            & 0.46          & 0.651                 & 0.2$\pm_{0.2}^{0.7}$      & [3]         \\                   
A3112                 & 75.2 &118    & 6$\pm 2$                 &$<$8.02          & 9$\pm3$            & 6.75          & 9.55                  & 0.8$\pm_{0.6}^{1.8}$      & [2]         \\                   
Centaurus             & 11.7 &26.2   & 3.3$\pm 0.1$             &5.5$\pm0.3 $     & 0.6$\pm0.1$        & 1.71          & 2.42                  & 0.15$\pm_{0.04}^{0.5}$    & [7]         \\                                
A3581                 & 18.8 &20.3   & 3.0$\pm 0.4$             &5$\pm2 $         & 2.1$\pm0.9$        & 2.28          & 3.23                  & 0.6$\pm_{0.4}^{1.7}$      & [2]         \\         
A4038                 & 16.7 &40.1   & 2.0$\pm 0.5$             &$<$2.39          & 3.1$\pm0.9$        & /             & /                     & /                         &             \\                     
A4059                 & 10.8 &36.5   & 2.0$\pm 0.5  $           &4$\pm2 $         & $<$1.51            & 3.90          & 5.52                  & 0.3$\pm_{0.2}^{1.0}$      & [2]         \\                     
AS1101                & 156  &227    & $<$3.80                  &$<$5.90          & $<$5.09            & 8.23          & 11.6                  & 0.9$\pm_{0.6}^{1.5}$      & [6]         \\                        
AWM7                  & 3.79 &27.7   & 2.1$\pm 0.3 $            &2.0$\pm0.7 $     & 2.2$\pm0.6$        & /             & /                     & 0.3$\pm_{0.1}^{0.3}$      &             \\                     
EXO0422-086           & 26.6 &38.6   & $<$1.53                  &$<$4.28          & $<$2.61            & /             & /                     & /                         &             \\                     
Fornax                & /    &/      & 0.04$\pm 0.01$           &0.85$\pm0.05 $   & $<$0.01            & /             & /                     & /                         &             \\                     
Hydra A               & 86.4 &109    & $<$5.25                  &$<$1.93          & $<$10.8            & 0.272         & 0.385                 & 4$\pm_{2}^{7}$            & [2]         \\                     
Virgo                 & 11.1 &37.4   & 0.62$\pm 0.03$           &2.33$\pm0.09 $   & $<$0.07            & 0.46          & 0.651                 & 0.1$\pm_{0.1}^{1.8}$      & [4]         \\           
MKW3s                 & 24.9 &48.2   & 3$\pm 1$                 &$<$2.10          & 5$\pm2$            & 1.33          & 1.88                  & 0.3$\pm_{0.2}^{0.4}$      & [4]         \\                   
MKW4                  & 4.55 &7.24   & 0.11$\pm 0.06$           &0.9$\pm0.2 $     & $<$0.06            & /             & /                     & /                         &             \\ \hline          
HCG62                 & 3.66 &4.46   & 1.5$\pm 0.2$             &/                &/                   & 0.0827        & 0.117                 & 0.06$\pm_{0.06}^{1.05}$   & [2]         \\             
NGC5044               & 13.5 &34.6   & $<$0.10                  &/                &/                   & 0.513         & 0.726                 & 0.20$\pm_{0.05}^{0.06}$   & [2]         \\           
NGC5813               & 3.34 &4.8    & $<$0.08                  &/                &/                   & 0.0414        & 0.0586                & 0.04$\pm_{0.01}^{0.01}$   & [2]         \\                                          
NGC5846               & 2.22 &4.3    & 0.43$\pm 0.06$           &/                &/                   & 0.0722        & 0.10                  & 0.09$\pm_{0.04}^{0.06}$   & [2]         \\ \hline

\end{tabular}}
% \end{center}
$\dot M_{\rm simple, 3Gyr}$ and $\dot M_{\rm simple, 7.7Gyr}$ are the `simple' and classical cooling rates in the absence of heating, which are deduced from \citet{2009ApJS..182...12C}. 
The measured cooling rates assume isobaric cooling flows (see equation \ref{equ:0}), 
where $\dot M_{1 cie + 1 cf}$ is measured from the \textit{cie} temperature $T_{\rm Max, \,1 cie + 1 cf}$ down to 0.01 keV in the one-stage model (see Table \ref{table:2cf_2}),
$\dot M_{\rm H, \,1 cie + 2 cf}$ is the cooling rate of the hotter cooling flow component from $T_{\rm Max, \,1 cie + 2 cf}$ down to 0.7 keV 
and $\dot M_{\rm C, \,1 cie + 2 cf}$ is measured between 0.7 and 0.01 keV both in the two-stage model. 
$L_{\rm H \alpha}$ is expressed in $10^{40}\,\rm erg\, \rm s^{-1}$ and converted into $\dot M_{\rm neb}$ using equation \ref{equ:2}.
The references for $L_{\rm H \alpha}$ are  
[1] \citet{2007AJ....134...14D},
[2] \citet{2016MNRAS.460.1758H},
[3] \citet{1999MNRAS.306..857C},
[4] \citet{1989ApJ...338...48H},
[5] \citet{2006MNRAS.371...93W},
[6] \citet{2005MNRAS.360..748J},
[7] \citet{2005MNRAS.363..216C}.
We use the star formation rates from \citet{2018ApJ...858...45M}.
All of the mass rates are measured in $\Msunpyr$.
\end{table*}

\begin{table*}
\caption{Key parameters continued.}  
 \vspace{-0.25cm}
\label{table:2cf_2}
\renewcommand{\arraystretch}{1.1}
 
\scalebox{0.95}{%
\hspace*{-1cm}\begin{tabular}{c c c c c c c c c}     
\hline\hline            
Source            & $T_{\rm Max, \,1 cie + 1 cf}$ & $T_{\rm Max, \,1 cie + 2 cf}$    & Source      &  $T_{\rm Max, \,1 cie + 1 cf}$  & $T_{\rm Max, \,1 cie + 2 cf}$\\ \hline                                                                                                    
2A0335+096        & 1.66$\pm 0.03$          & 1.81$\pm 0.05$           & Centaurus         & 1.52$\pm 0.01$ & 1.82$\pm 0.05$   \\ 
A85               & 3.14$\pm 0.13$          & 3.18$\pm 0.15$           & A3581             & 1.40$\pm 0.02$ & 1.43$\pm 0.03$   \\ 
A133              & 2.36$\pm 0.10$          & 2.54$\pm 0.14$           & A4038             & 2.46$\pm 0.16$ & 2.42$\pm 0.15$   \\ 
A262              & 1.48$\pm 0.01$          & 1.60$\pm 0.04$           & A4059             & 2.57$\pm 0.12$ & 2.64$\pm 0.14$   \\ 
Perseus 90$\%$ PSF& 2.36$\pm 0.06$          & 2.21$\pm 0.04$           & AS1101            & 1.96$\pm 0.05$ & 1.96$\pm 0.04$   \\ 
Perseus 99$\%$ PSF& 2.14$\pm 0.02$          & 2.05$\pm 0.02$           & AWM7              & 2.02$\pm 0.12$ & 2.02$\pm 0.13$   \\ 
A496              & 2.20$\pm 0.05$          & 2.27$\pm 0.08$           & EXO0422-086       & 2.32$\pm 0.19$ & 2.30$\pm 0.17$   \\ 
A1795             & 3.10$\pm 0.16$          & 3.12$\pm 0.15$           & Fornax            & 0.99$\pm 0.01$ & 1.56$\pm 0.04$   \\ 
A1835             & 4.34$\pm_{0.32}^{0.22}$ & 4.36$\pm_{0.33}^{0.25}$  & Hydra A           & 2.44$\pm 0.08$ & 2.43$\pm 0.09$   \\ 
A1991             & 1.71$\pm 0.08$          & 1.88$\pm 0.14$           & Virgo             & 1.37$\pm 0.01$ & 1.43$\pm 0.01$   \\ 
A2029             & 3.47$\pm 0.14$          & 3.46$\pm 0.13$           & MKW3s             & 2.34$\pm 0.08$ & 2.33$\pm 0.08$   \\ 
A2052             & 1.87$\pm 0.05$          & 2.07$\pm 0.07$           & MKW4              & 1.46$\pm 0.01$ & 1.53$\pm 0.03$   \\ 
A2199             & 2.73$\pm 0.11$          & 2.76$\pm 0.11$           & HCG62             & 0.90$\pm 0.01$ &  /               \\ 
A2597             & 2.62$\pm 0.11$          & 2.67$\pm 0.11$           & NGC5044           & 0.87$\pm 0.01$ &  /               \\ 
A2626             & 3.22$\pm_{0.37}^{0.54}$ & 3.38$\pm_{0.47}^{0.62}$  & NGC5813           & 0.68$\pm 0.01$ &  /               \\                                                                                                                                                                             
A3112             & 2.66$\pm 0.09$          & 2.64$\pm 0.09$           & NGC5846           & 0.75$\pm 0.01$ &  /               \\ \hline                
\end{tabular}}
\\The temperatures of the \textit{cie} component (the maximum temperature of the cooling flow) in both the one-stage and two-stage models.
\end{table*}

\begin{table}
\caption{One-stage model for galaxies.}  
 \vspace{-0.25cm}
\label{table:1cf}
\renewcommand{\arraystretch}{1.1}
 
\scalebox{0.8}{%
\hspace*{-0.75cm}\begin{tabular}{c c c c c c}     
\hline\hline            
Source  &  $\dot M_{\rm 1 cie + 1 cf}$ & $T_{\rm Max, \,1 cie + 1 cf}$ & Source  &  $\dot M_{\rm 1 cie + 1 cf}$ & $T_{\rm Max, \,1 cie + 1 cf}$    \\ \hline                                                                                                    
M49     & 0.07$\pm 0.02$ & 0.92$\pm 0.01$ & NGC1550 & 0.5$\pm 0.1$   & 1.17$\pm 0.01$\\
M86     & 0.20$\pm 0.08$ & 0.84$\pm 0.02$ & NGC3411 & $<$0.32        & 0.91$\pm 0.01$\\
M89     & 0.3$\pm 2$     & 0.64$\pm 0.02$ & NGC4261 & $<$0.10        & 0.73$\pm 0.01$\\
NGC507  & 0.69$\pm 0.06$ & 1.17$\pm 0.01$ & NGC4325 & $<$2.67        & 0.90$\pm 0.02$\\
NGC533  & $<$2.64        & 0.90$\pm 0.04$ & NGC4374 & 0.2$\pm 0.1$   & 0.71$\pm 0.04$\\
NGC1316 & 0.26$\pm 0.04$ & 1.34$\pm 0.03$ & NGC4636 & 1.1$\pm 0.2$   & 0.72$\pm 0.01$\\
NGC1404 & 0.6$\pm 0.3$   & 0.69$\pm 0.03$ & NGC4649 & $<$0.01        & 0.84$\pm 0.01$\\\hline               
\end{tabular}}
\\Only the one-stage cooling flow model is used for galaxies.
\end{table}

The one-stage cooling flow model includes 1 \textit{cie} and 1 \textit{cf} components which have the same abundances. 
We assume that the maximum temperature of the cooling flow is the same as the \textit{cie} temperature, and the minimum temperature is fixed at 0.01 keV. 

We find a low level of cooling rate in most clusters, typically less than $10\,\Msunpyr$ (see Table \ref{table:2cf}). 
For clusters with \textit{cie} temperature higher than 1.6 keV, the measured cooling rates are compared with the `simple' cooling rates $\dot M_{\rm simple}(<r)= M_{\rm gas}(r)/t_{\rm cool}(r)$, 
where $M_{\rm gas}(r)$ is the total gas mass enclosed within a radius $r$. 
The radius is determined where the radiative cooling time $t_{\rm cool}(r)$ is 3 Gyr.
We use the electron density and the cooling time profiles in \citet{2009ApJS..182...12C}, where we model the electron density by a power law with the function of form $n_{e}=Ar^{b}$. 
We assume the clusters are spherically symmetric and integrate the electron densities between 0.1 kpc and the radius where the cooling time is 3 Gyr. 
This calculation will give approximately 20$\%$ systematic uncertainty from the actual density profile and the asymmetry of clusters. 
As an example, if we use the density profiles of the eastern and western halves of the Centaurus cluster (\citealt{2016MNRAS.457...82S}), 
the `simple' cooling rate is slightly lower at $9.5\,\Msunpyr$, as oppose to $11.7\,\Msunpyr$ from the symmetric density profile. 
These `simple' cooling rates are consistent with the clusters values calculated by \citealt{2018ApJ...858...45M}. 
We repeat the same calculation for the classical cooling rates, which have the radiative cooling time of 7.7 Gyr. 
In this work, we denote the `simple' and classical cooling rates as $\dot M_{\rm simple, 3Gyr}$ and $\dot M_{\rm simple, 7.7Gyr}$ respectively
\footnote{Note that the classical cooling rate differs from that determined if a cooling \textit{flow} has been established. 
If the gas flows inward, gravitational energy is released and must be accounted for (\citealt{1985MNRAS.216..923F}).}. 
In general, we find that $\dot M_{\rm simple, 3Gyr}<\dot M_{\rm simple, 7.7Gyr}$. 

Since the `simple' and classical cooling rates can serve as a proxy for predicted cooling rates in the absence of heating, 
we can infer the efficiency of heating due to feedback between the measured-to-predicted ratio and unity. 
The distribution is shown in the top panel of Fig. \ref{fig:r_hist} and \ref{fig:r_hist_2}.
The great majority of clusters have a measured-to-predicted ratio less than 0.4 if we use $\dot M_{\rm simple, 3 Gyr}$, which is equivalent to a minimum heating efficiency of 60$\%$. 
19 out of 22 clusters have a measured-to-predicted ratio less than 0.2.
In the $\dot M_{\rm simple, 7.7 Gyr}$ case, the measured cooling rates are less than 30$\%$ of $\dot M_{\rm simple, 7.7 Gyr}$ for all clusters, 
and less than 10$\%$ in 18 out of 22 clusters, which is consistent with \citet{2010A&A...513A..37H}. 
We further notice that clusters with upper limits in the measured cooling rates are suppressed more effectively than those with measurements. 

For groups, very weak cooling flows are sometimes detected, typically less than $1\,\Msunpyr$, and many objects only have upper limits in the measured cooling rates (see Table \ref{table:1cf}). 
The level of cooling rates is similar to the values reported by \citet{2005ApJ...635.1031B}, usually consistent within 1 $\sigma$ uncertainty. 
The minimum temperature of 0.01 keV is important for groups since their \textit{cie} temperature is typically less than 1 keV.  
If we use a higher value, it is likely to overpredict the cooling rates when the range between the maximum and minimum temperatures is very narrow.  
Note that it is possible that some objects which can be well fitted by the 1 \textit{cie} model may also be fitted by a single cooling flow model (see e.g. \citealt{2014A&A...572L...8P}). 

\subsubsection{Two-stage cooling flow model}

\begin{figure*}
  \includegraphics[width=\columnwidth, angle=0]{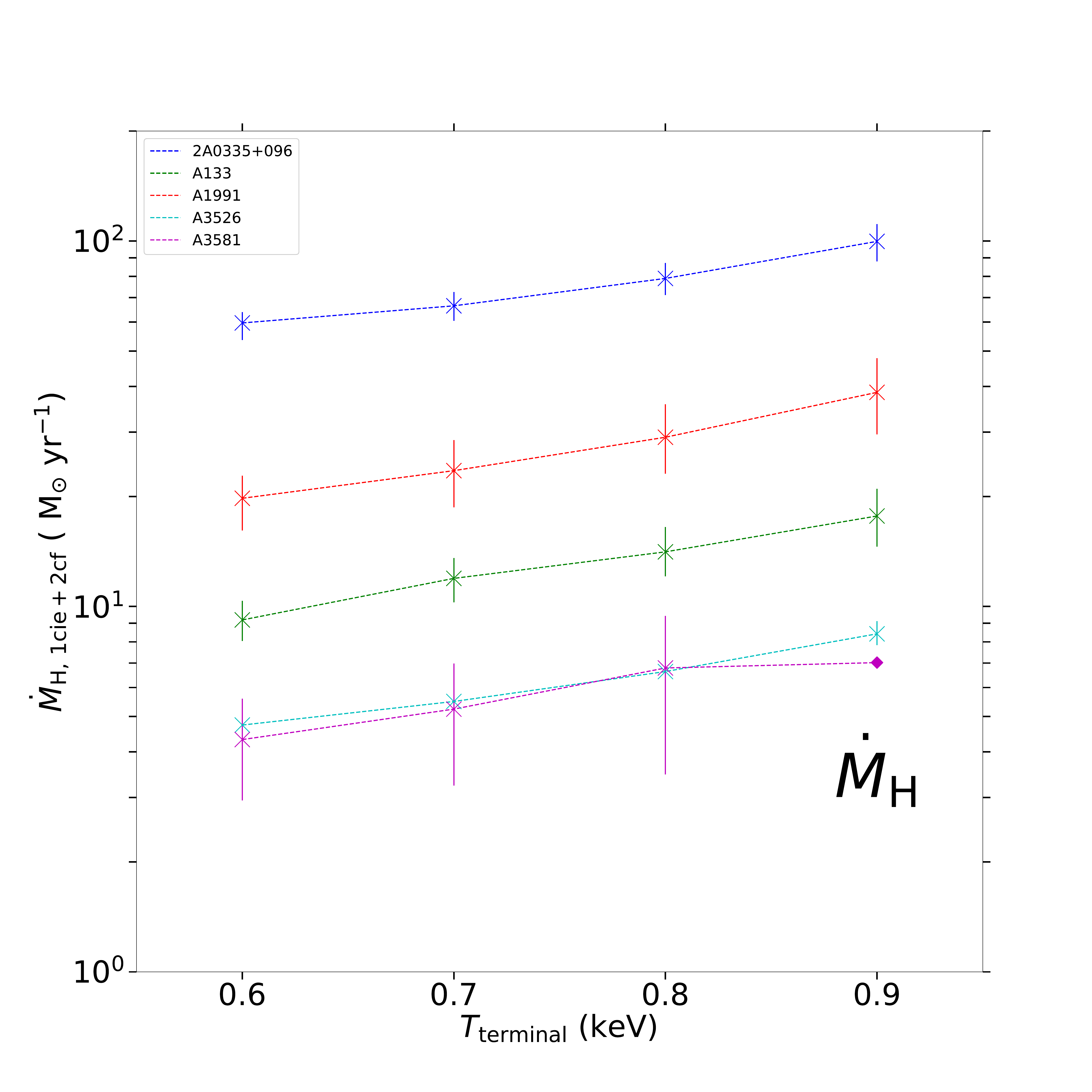}
  \includegraphics[width=\columnwidth, angle=0]{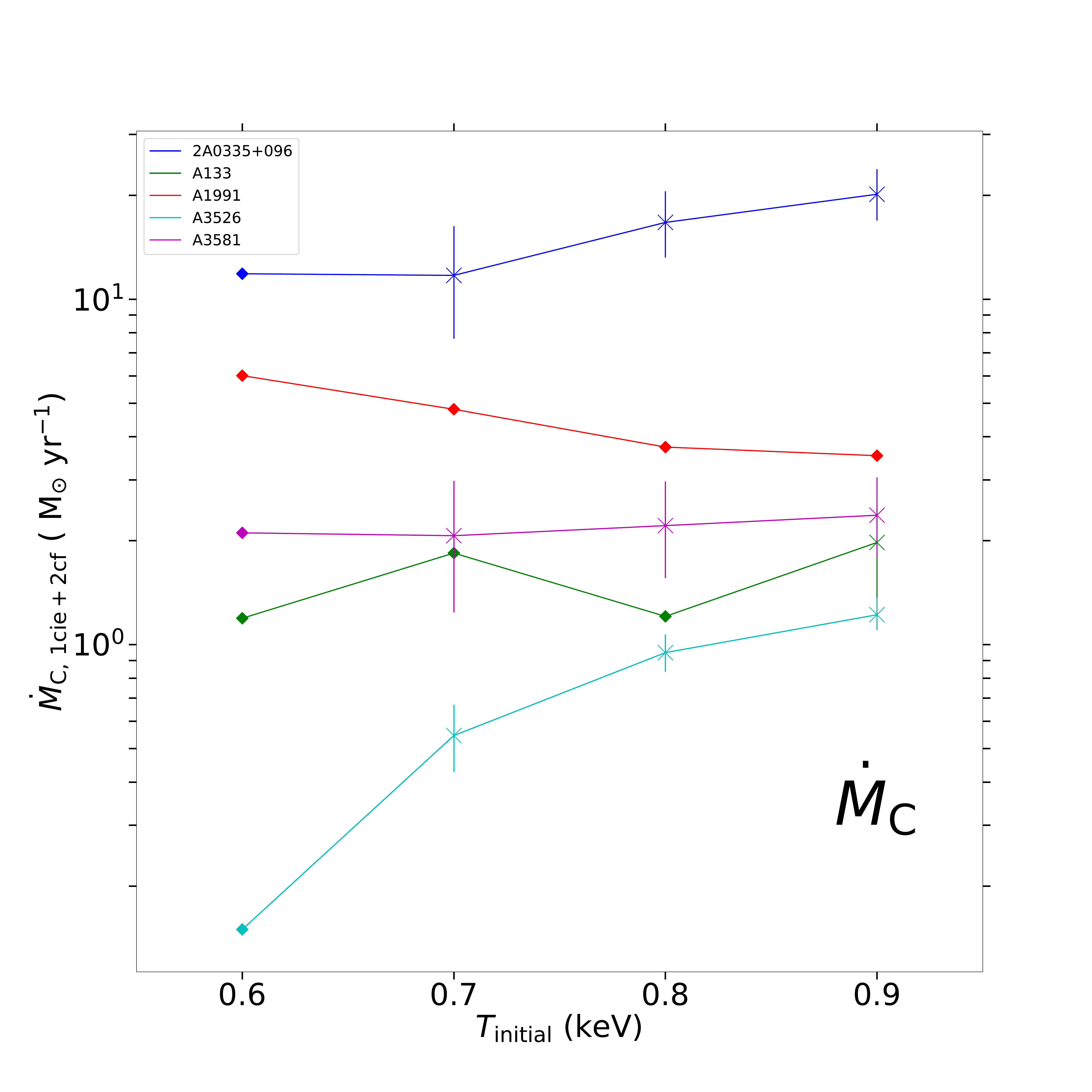}
  \caption{Left: (a) The measured cooling rates between the \textit{cie} temperature and different terminal temperatures in the two-stage cooling flow model. 
  The diamond shape represents 2 $\sigma$ upper limit.
  Right: (b) The residual cooling rates between different initial temperatures and 0.01 keV in the two-stage cooling flow model. 
  The initial temperatures match the terminal temperatures of the hotter cooling flow component.
  }
  \label{fig:mdot_temp}
\end{figure*}

For clusters, it is also possible that the cooling flow terminates at a temperature higher than 0.01 keV, 
because no {O\,\scriptsize{VII}} is seen in most objects (\citealt{2016MNRAS.461.2077P}). 
We choose a terminal temperature (i.e. the minimum temperature of the \textit{cf} component) of 0.7 keV, 
and use an additional \textit{cf} component to measure the residual cooling rate between 0.7 and 0.01 keV (two-stage model).
This terminal temperature is the average cooler temperature of all objects in the 2 \textit{cie} model, 
which is higher than the average terminal temperature of 0.6 keV if we set the terminal temperature free in the one-stage model. 
For 5 clusters, we show the measured cooling rates with different terminal temperatures $T_{\rm terminal}$ (Fig. \ref{fig:mdot_temp}a), 
and hence the initial temperatures $T_{\rm initial}$ for the residual cooling rates (Fig. \ref{fig:mdot_temp}b). 
It is seen that the cooling rates of the hotter cooling flow component gradually decrease with lower terminal temperatures, and do not change our conclusions. 
The trend is slightly more complicated for the residual cooling rates, where more upper limits are detected. 
They do not always decrease as the cooling rates of the hotter cooling flow, because fitting the component in a narrower temperature range may boost the measured values. 
Additionally, we simulate the spectrum of a cooling flow of $100\,\Msunpyr$ between 0.01 and 5 keV using the response matrix of A133, 
which is then fitted by two cooling flow components with the terminal temperature of the hotter cooling flow component changing between 0.5 and 0.9 keV, 
and find that both $\dot M_{\rm H, \,1 cie + 2 cf}$ and $\dot M_{\rm C, \,1 cie + 2 cf}$ are consistent with $100\,\Msunpyr$ within 1 $\sigma$ uncertainty.
Therefore, we only present our measured cooling rates with a terminal temperature of 0.7 keV in Table \ref{table:2cf}. 

\begin{figure}
  \includegraphics[width=0.95\linewidth, angle=0]{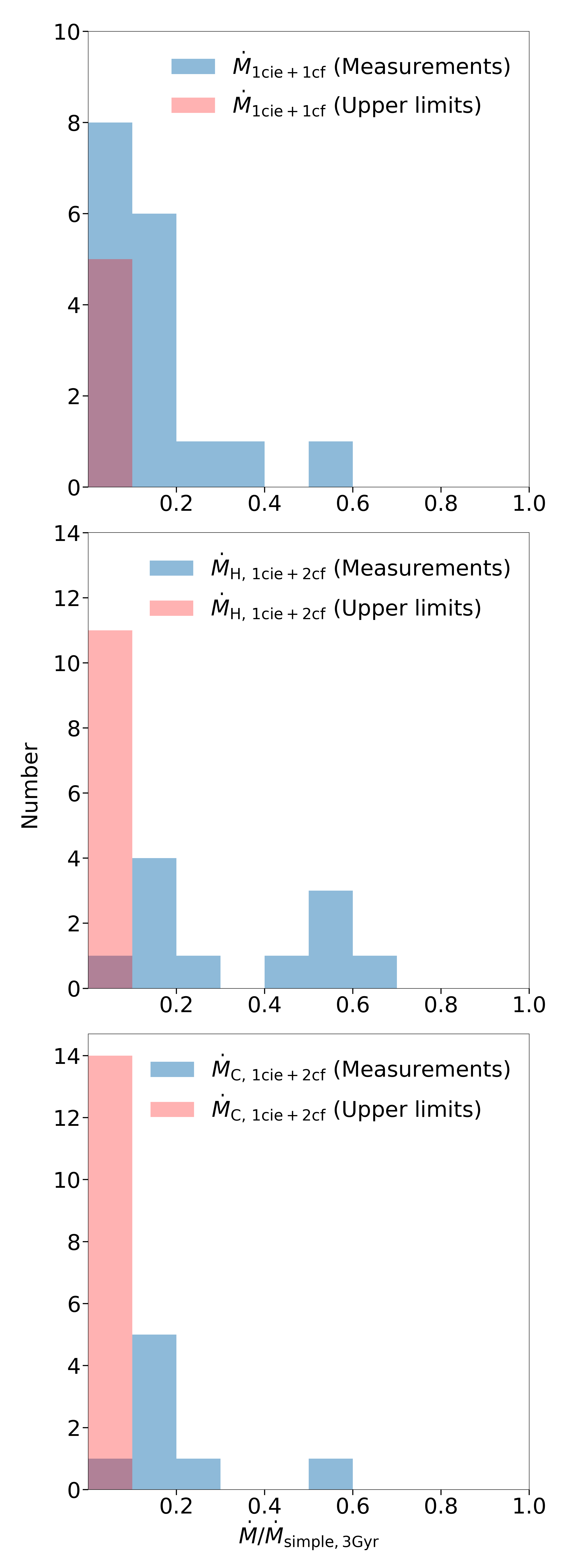}
  \caption{The histograms of the ratio of measured cooling rates to $\dot M_{\rm simple, 3 Gyr}$ for clusters with $T_{\rm Max, \,1 cie + 2 cf}$ $>$ 1.6 keV, 
  where $\dot M_{\rm simple, 3 Gyr}$ are deduced from the gas parameters tabulated by \citet{2009ApJS..182...12C}. 
  Top: cooling rates of the one-stage model. 
  Middle: cooling rates between $T_{\rm Max, \,1 cie + 2 cf}$ and 0.7 keV in the two-stage model. 
  Bottom: the residual cooling rates below 0.7 keV.
  }
  \label{fig:r_hist}
\end{figure}

\begin{figure}
  \includegraphics[width=0.95\linewidth, angle=0]{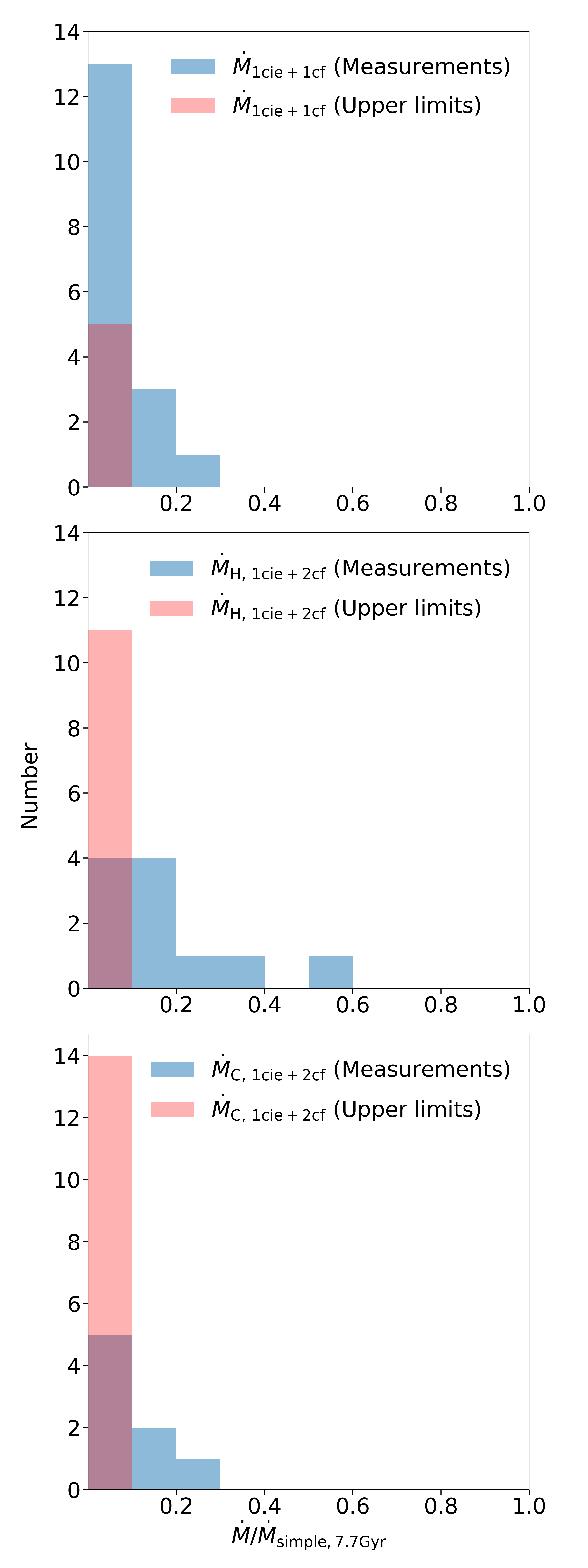}
  \caption{As Fig. \ref{fig:r_hist}, but the measured cooling rates are now compared to $\dot M_{\rm simple, 7.7 Gyr}$ (\citealt{2009ApJS..182...12C}).
  }
  \label{fig:r_hist_2}
\end{figure}

The two-stage cooling flow model has one more free parameter than the one-stage model, which means more degeneracy during spectral fitting. 
Therefore, we observe more clusters with their cooling rates as upper limits, which has a significant impact on e.g. Perseus and A1835.
In general, the cooling rates of hotter \textit{cf} component are either higher than the residual cooling rates, or consistent within 1 $\sigma$ uncertainty. 
Both the cooling and residual cooling rates are compared with $\dot M_{\rm simple}$ in Fig. \ref{fig:r_hist} and \ref{fig:r_hist_2}. 
It is seen that the cooling rates above 0.7 keV are suppressed by at least 30$\%$ in most objects if we use $\dot M_{\rm simple, 3 Gyr}$, except A262 which has a measured-to-predicted greater than unity. 
If we instead use $\dot M_{\rm simple, 7.7 Gyr}$, this ratio becomes 40$\%$. 
The residual cooling rates are much lower and therefore the ratios seen in the bottom panels are much lower. 
For approximately 90$\%$ of clusters, the residual cooling rates are suppressed by at least 80$\%$. 
Therefore, the two-stage cooling flow model suggests that clusters seem to cool only down to 0.7 keV. 
The \textit{cie} temperature of both the one-stage and two-stage cooling flow models are listed in Table \ref{table:2cf_2}.

\subsubsection{Comparing the cooling flow models}

Comparing the two cooling flow models, we find that the cooling rates of the hotter \textit{cf} component of the two-stage model are generally higher than the one-stage cooling rates. 
This is because the hotter \textit{cf} component of the two-stage model need to contribute to {Fe\,\scriptsize{XVII}} emission between a narrow temperature range (0.7-0.9 keV) where its emissivity dominates, 
and the one-stage model can contribute to {Fe\,\scriptsize{XVII}} emission between 0.01-0.9 keV. 
Since the two-stage cooling flow model is fitting one more parameter than the one-stage model, we are also interested in whether the two-stage model is statistically better and the difference in spectral fit.

We find that the two-stage model has a lower C-stat than the one-stage model ($\Delta$C-stat$\geq 10$ for 1 degree of freedom) in 9 out of 22 clusters (see Table. \ref{table:chi2}). 
However, there are a few special clusters where we prefer the one-stage model, such as Perseus and MKW3s, because the residual cooling rate is higher than the cooling rate above 0.7 keV. 
For AWM7, we find that there is no difference between the two cooling flow models, both in terms of C-stat and cooling rates. 
Therefore, we conclude that the one-stage cooling flow is sufficient for AWM7
\footnote{AWM7 is unusual: Chandra data (\citealt{2012MNRAS.421..726S}) shows a small ($\sim10$ kpc) bright core with a low cooling time in a hotter diffuse medium.}.

\begin{figure}
  \includegraphics[width=1.8\linewidth, angle=0]{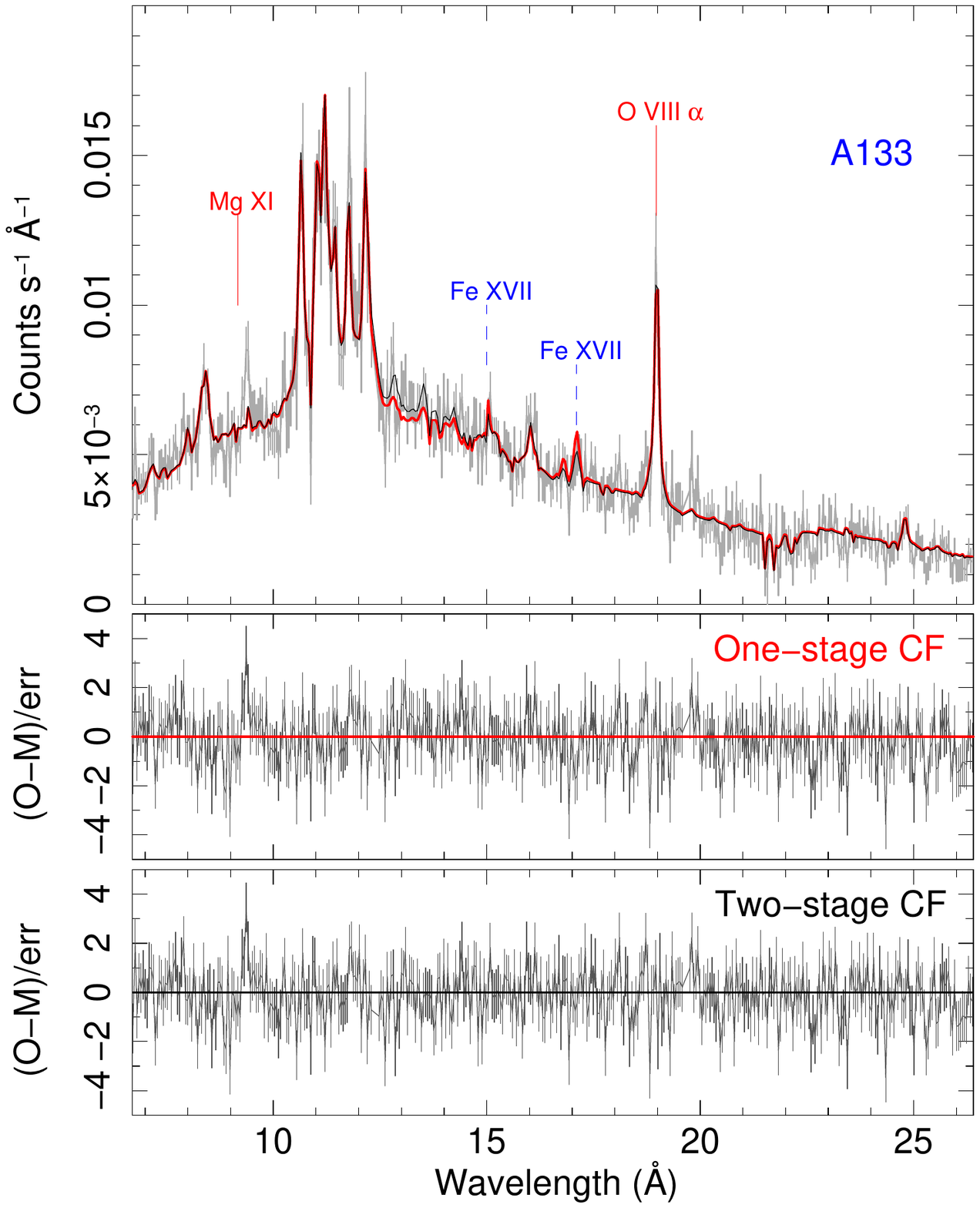}
  \caption{The spectrum (grey) and best fit cooling flow models of A133 in rest wavelength. 
  The spectrum has a peak at around 9.4 \AA, which is an artefact and cannot be fitted. 
  }
  \label{fig:A133}
\end{figure}

\begin{figure}
  \includegraphics[width=1.8\linewidth, angle=0]{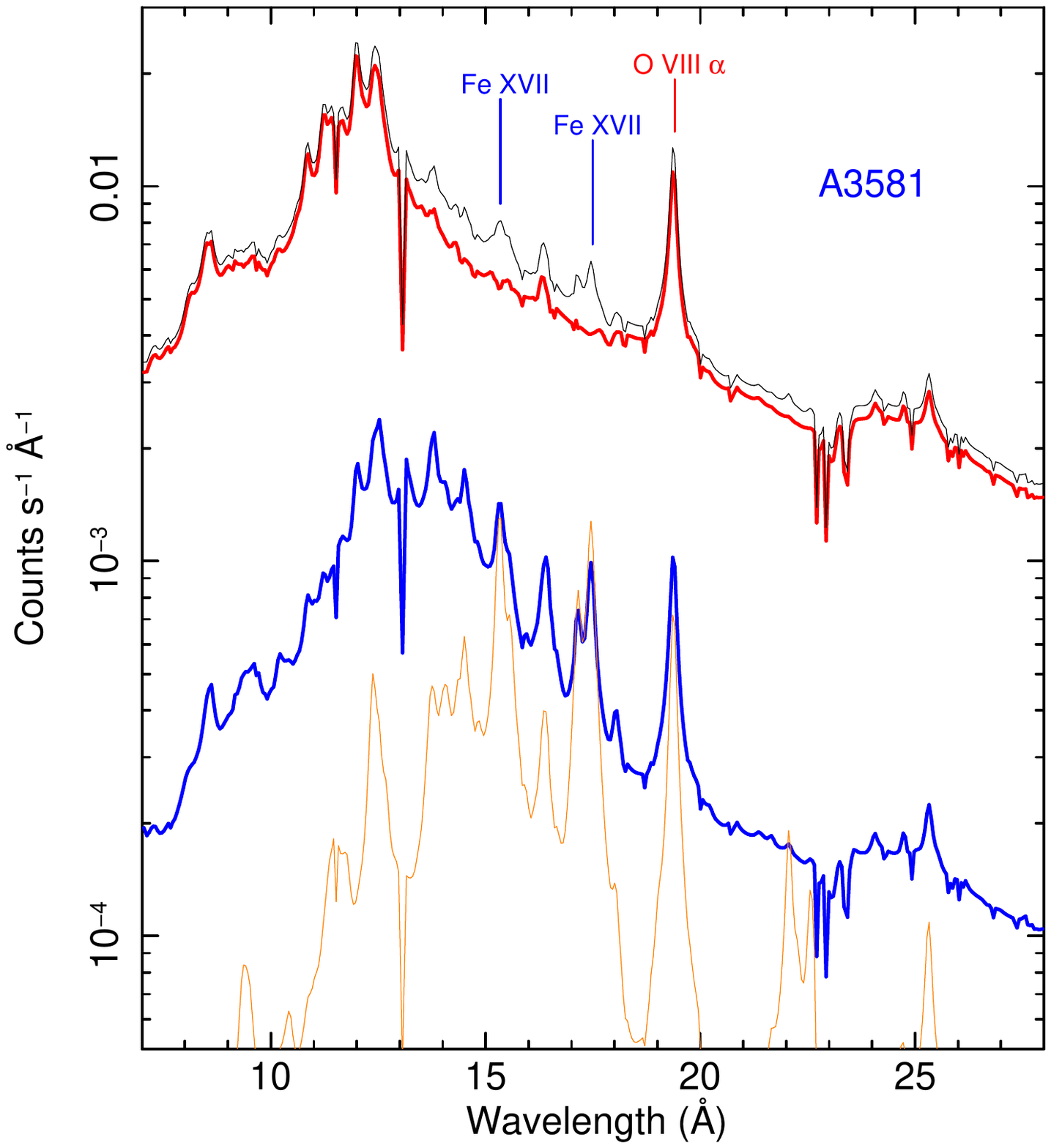}
  \caption{The best fit two-stage cooling flow model of A3581. 
  The top black thin line is the total emission, 
  the thick red line is the contribution from the \textit{cie} component, 
  the thick blue line is the hotter cooling flow component and the thin orange line at the bottom is the residual cooling flow.
  }
  \label{fig:A3581}
\end{figure}

To demonstrate the difference in the spectral fit, we show the spectrum and cooling flow models of A133 in Fig. \ref{fig:A133}. 
The two-stage improves the spectral fit to the {Fe\,\scriptsize{XVII}} forbidden line at 17.1 \AA, which is related to the cooling gas. 
Furthermore, the models differ between 12.5 and 14.5 \AA, which has emission from different ionisation stages of Fe which peak at hotter temperatures, e.g. {Fe\,\scriptsize{XX}}. 
However, neither of cooling flow models is significantly better at these wavelengths. 
We also show the contribution from different components in the two-stage model in Fig. \ref{fig:A3581}. 
It is seen that the contribution from the two cooling flow components are comparable at important lines, e.g. {Fe\,\scriptsize{XVII}}, {O\,\scriptsize{VIII}} and {O\,\scriptsize{VII}}.
In conclusion, we have statistical and spectral evidence that the two-stage cooling flow model is better in at least 9 out of 22 clusters.

\subsection{Special clusters}
\label{sec:special}

\subsubsection{Perseus and Virgo}
\label{sec:Perseus}
It is well known that both the X-ray bright Perseus and Virgo clusters have a bright variable AGN at the centre, and it is well described by a \textit{pow} component.  
The X-ray emission of the AGN can vary by an order of magnitude in only a few years, and hence we need to fix the parameters of the \textit{pow} component at the time of our observation. 
\citet{2003ApJ...590..225C} found that the AGN emission in Perseus can be well fitted by an absorbed ($N_{\rm H} = 10^{21}\,\rm cm^{-2}$) \textit{pow} component with a photon index of 1.65, 
where such a column density is comparable to $N_{\rm HI}$ from our own galaxy. 
The luminosity of the nucleus (OBSID = 0085110101) is constrained to be of the order of $10^{43}\,\rm erg\,\rm s^{-1}$ in the 0.5-8 keV band with 20$\%$ systematic uncertainties.
In this work, we choose the emission measure such that the \textit{pow} component only produces $10^{43}\,\rm erg\,\rm s^{-1}$ in the same energy band. 
We fit both the 90 and 99$\%$ PSF (0.8$^{\prime}$ and 3.4$^{\prime}$) spectra in this work, and the variable AGN gives an additional 5 and 15 $\Msunpyr$ statistical uncertainty respectively in the one-stage cooling flow model. 
The actual cooling structure of Perseus is likely to be more complicated than our models because of the existence of the {O\,\scriptsize{VII}} emission (\citealt{2016MNRAS.461.2077P}), which is beyond the scope of this work.

The initial study on Virgo suggests that it is inadequate to use only the stacked spectrum, and hence we perform simultaneous spectral fitting on the non-stacked spectra. 
The difference in spectra between the two observations is purely due to the variation of the \textit{pow} component since the ICM emission is constant. 
We choose the photon index of the first observation to be 2.4 with a flux of $3.76\,\times\,10^{-12}\,\rm erg\,\rm cm^{-2}\,\rm s^{-1}$ in 0.3-8 keV 
(Observation date: \textit{Chandra}= 30 July 2000, XMM-\textit{Newton}= 19 June 2000; see \citealt{2004ApJ...617..915D}). 
For the second observation, we use the same photon index and set the emission measure free. 
The \textit{pow} component of the second observation yields a flux of $5.19\,\times\,10^{-11}\,\rm erg\,\rm cm^{-2}\,\rm s^{-1}$, 
which is 13.8 times brighter than the previous observation, and comparable to the ICM luminosity.

\subsubsection{Centaurus}
 
In addition to the multi-temperature and cooling flow models, we apply a 5 \textit{cf} model to both the 90 and 99 $\%$ PSF spectra of the Centaurus cluster, which has been used by \citet{2008MNRAS.385.1186S}. 
The temperatures of the \textit{cf} components are 3.2-2.4, 2.4-1.6, 1.6-0.8, 0.8-0.4, 0.4-0.0808 keV.
The components below 0.8 keV are convolved by the same \textit{lpro} component, and we use a second \textit{lpro} component for the hotter \textit{cf} components (3.2 to 0.8 keV) to improve the spectral fit. 
From the spatial analysis by \citet{2008MNRAS.385.1186S}, the components below 0.8 keV are located in the innermost core, which supports this critical temperature. 
The best fit values are shown in Table \ref{table:centaurus}. 

We compare these cooling rates of the 90$\%$ PSF spectrum with the two-stage model in section \ref{sec:cooling_flow}. 
The residual cooling rate $\dot M_{\rm C, \,1 cie + 2 cf}$ is lower than Comp 4 (0.8-0.4 keV), and Comp 5 only gives an upper limit. 
They suggest that the cluster cools below 0.7 keV, and stops cooling potentially at around 0.2 keV, where {O\,\scriptsize{VII}} emission was found by \citet{2016MNRAS.461.2077P}. 
However, it is difficult to constrain the exact terminal temperature, since the {O\,\scriptsize{VII}} lines are very weak comparing to the continuum. 

For the 99$\%$ PSF spectrum, we find our results different from those reported by \citet{2008MNRAS.385.1186S}. 
Our model does not fit the hottest \textit{cf} component (3.2-2.4 keV), which has unexpectedly low upper limits. 
Further investigation suggests that this can be strongly influenced by different free abundances and the calibration below 8.5\,\AA\, is very poor. 
Nevertheless, we can still confirm that the Centaurus cluster can be resolved with more than two components (see Table \ref{table:centaurus} for \textit{cf} models).

\begin{table}
\caption{The best fit parameters for the Centaurus cluster.}  
\vspace{-0.25cm}
\label{table:centaurus}      
\renewcommand{\arraystretch}{1.1}

 \small\addtolength{\tabcolsep}{+2pt}
 
\scalebox{.8}{%
\hspace*{-0.75cm}\begin{tabular}{c c c c c}     
\hline\hline            
               & $\dot M$ (90$\%$ PSF)& $\dot M$ (99$\%$ PSF) & $\dot M$ from \citet{2008MNRAS.385.1186S}\\ \hline                                                                                                                         
Comp 1         & $<$ 6.41             & $<$ 15.9              & 46.4$\pm 2.7$                            \\                     
Comp 2         & 21.2$\pm 2.43$       & 60.0$\pm 4.56$        & 32.1$\pm 2.8$                            \\                    
Comp 3         & 4.69$\pm 0.23$       & 7.06$\pm 0.30$        & 6.30$\pm 0.37$                           \\                    
Comp 4         & 0.99$\pm 0.08$       & 1.37$\pm 0.13$        & 2.13$\pm 0.48$                           \\                    
Comp 5         & $<$ 0.13             & $<$ 0.16              & $<$ 2.25                                 \\ \hline                
\end{tabular}}

The temperature grids are 3.2-2.4, 2.4-1.6, 1.6-0.8, 0.8-0.4, 0.4-0.0808 keV (from Comp 1 to Comp 5). 
The cooling rates are in $\Msunpyr$. 
The second and third columns are our measured values.
The fourth column has been revised using the update XSPEC package, where the upper limit in Comp 5 is 1 $\sigma$ only. 

\end{table}

Unfortunately, 5-component models cannot be repeated for other objects. 
Since the thermal structures of clusters are intrinsically different, it is very difficult to define a universal temperature grid.
The limited statistics in many objects also forbid us from resolving them further.
Hence, we have limited a maximum of 2 \textit{cf} components for the remaining clusters. 

\section{Discussion}
\label{sec:discussion}

\subsection{The missing soft X-ray emission and H$\alpha$ filaments}

Our spectral fits indicate that little cooled gas is seen in the X-ray band below 0.4 keV (Table \ref{table:2cie}), 
and the upper limits obtained on $\dot M_{\rm C, \,1 cie + 2 cf}$ show little evidence for cooling gas either. 
Gas may be cooling down due to X-ray emission above $\sim0.7$\,keV but if it does then it does not continue cooling in that way below that energy.
Such behaviour is peculiar because the radiative cooling time shortens rapidly as the gas temperature drops.

The situation was modelled by \citet{2003MNRAS.344L..48F} in terms of `missing' soft X-ray luminosity, $L_{\rm miss}$. 
This is the luminosity difference between gas cooling at a rate $\dot M$ down to a terminal temperature and gas cooling at the same rate to zero. 
It was noted that $L_{\rm miss}$ is similar to the luminosity of the optical-UV emission-line nebulosity $L_{\rm neb}$ around the Brightest Cluster Galaxies (BCGs) of the cool core clusters studied, 
suggesting that $L_{\rm neb}$ was powered by the remaining thermal energy of the cooling gas at $T_{\rm min}$. 
The hypothesis is supported by the spatial-coincidence of the nebulae and the soft X-ray emission 
(e.g. Perseus: \citealt{2003MNRAS.344L..48F,2006MNRAS.366..417F}; Centaurus: \citealt{2005MNRAS.363..216C}, \citealt{2016MNRAS.461..922F}; A1795: \citealt{2001MNRAS.321L..33F}, \citealt{2005MNRAS.361...17C}). 
In order to capture the full emission from the nebula, we increase the observed H$\alpha$ luminosity by a factor of 15
\footnote{The actual ratio of $L_{\rm neb}$ to $L_{\rm H\alpha}$ should be 10 to 20.} (\citealt{2009MNRAS.392.1475F}). 

\begin{figure*} \includegraphics[width=0.9\linewidth,
angle=0]{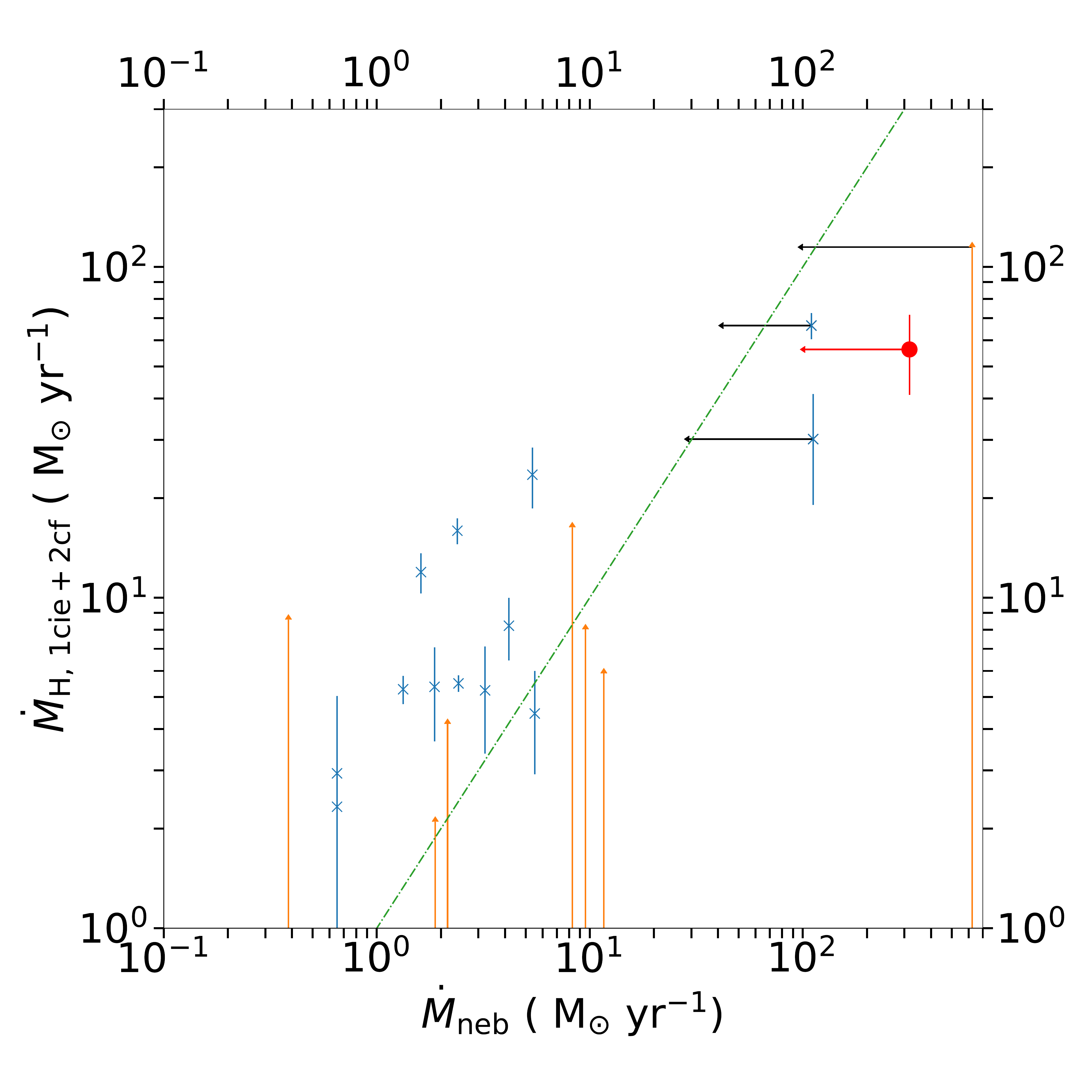}
  \caption{The cooling rates $\dot M_{\rm H, \,1 cie + 2 cf}$ in the two-stage model (between $T_{\rm Max, \,1 cie + 2 cf}$  and 0.7 keV) against
thermal energy required to sustain emission at longer wavelengths in filaments. 
The horizontal arrows give the extreme values of $\dot M_{\rm neb}$ when using $T_{\rm Max, \,1 cie + 2 cf}$ instead of 0.7 keV in equation \ref{equ:2}. 
For the Perseus cluster, the cooling rate is measured from the 99$\%$ PSF spectrum in the one-stage model (red circle).}
  \label{fig:mass_compare_1}
\end{figure*}

The idea agrees with the heating and excitation of the cold gas being due to fast particles (\citealt{2009MNRAS.392.1475F}). 
\citet{2011MNRAS.417..172F} suggested that the fast particles are the result of interpenetration of the hot and cold gas. 
In this case, $L_{\rm miss}={3\over 2}\dot M_{\rm H, \,1 cie + 2 cf}{{kT}\over{\mu m_{\rm p}}}$
should equal $L_{\rm neb}= 15L_{\rm H\alpha}$.
We can rewrite the above as $\dot M_{\rm H, \,1 cie + 2 cf}= \dot M_{\rm neb}$,
where 
\begin{equation}
\label{equ:2}
\dot M_{\rm neb}= 0.99 \times (\frac{L_{\rm H \alpha}}{10^{40}\,\rm erg\,\rm s^{-1}})(\frac{kT}{1keV})^{-1}\rm M_{\odot}\,\rm yr^{-1},
\end{equation}
which is the mass inflow rate into the nebula of particles at energy 0.7 keV.
We plot $\dot M_{\rm H, \,1 cie + 2 cf}$ against $\dot M_{\rm neb}$ in Fig. \ref{fig:mass_compare_1}
\footnote{We caution against drawing any correlations from this and the next two plots since both axes involve the square of the distance.}. 
Most of the objects lying to the lower left in the plot and $\dot M_{\rm H, \,1 cie + 2 cf}>\dot M_{\rm neb}$, 
whereas the opposite is true for the 4 objects up and to the right of the plot. 

The situation is more complex than assumed above and we now examine a cluster from the right hand side (Perseus) and the left hand side (Centaurus) in more detail. 
The Perseus cluster has an extensive H$\alpha$ nebula (\citealt{2001AJ....122.2281C}; \citealt{2008Natur.454..968F}), 
and its H$\alpha$ luminosity is significantly higher than most other objects in our sample.
\citet{2003MNRAS.344L..48F} used the 4$^{\prime}$ region image of the Perseus cluster and found the filaments are UV/optically bright, 
where the soft X-ray emission is an energetically minor component. 
Therefore, we use a broader 3.4$^{\prime}$ spectrum for Perseus (red circle) here to measure the cooling rate. 
The two-stage cooling model is affected by resonance scattering, and the residual cooling rate below 0.7 keV is stronger than the cooling rate above. 
This is problematic because we expect residual cooling to be replenished by the cooling flow above 0.7 keV. 
Instead, we use the cooling rate $55.5 \pm\,15.3 \rm M_{\odot}\,\rm yr^{-1}$ in the one-stage cooling model, 
which is 30$\%$ lower than the cooling rate of a one-stage model that cools down to 0.7 keV. 
To better match this cooling rate to $\dot M_{\rm neb}$ we need to use gas at a higher temperature,
$\sim T_{\rm Max, \,1 cie + 1 cf}$, rather than 0.7\,keV. 
This would not show up in our analysis here and remains a possible, but unconfirmed solution for powering of the filaments. 
A detailed study with Chandra of the X-ray spectra of the nebula filaments in the Perseus cluster by \citet{2015MNRAS.453.2480W} reveals components at both $\sim2$\,keV and 0.7\,keV. 
Several other clusters (M87, Centaurus and A1795) also show the need for the 0.7\,keV component. 
If the Perseus nebula is powered by particles from the surrounding hot gas then an inflow from the surrounding gas at a rate of $\sim100 \Msunpyr$ is required.

For the Centaurus cluster (the blue point in Fig. \ref{fig:mass_compare_1}), to power the nebula requires an inflow of $\sim2 \Msunpyr$, 
which is significantly less than our RGS measured value of $\dot M_{\rm H, \,1 cie + 2 cf}= 5.2 \pm 0.2\Msunpyr$. 
However, we note that Chandra images of the soft X-ray emission in the cluster reveal emission which is much more extended than the bright filamentary nebula (\citealt{2016MNRAS.457...82S}). 
If we refine the estimate of $\dot M_{\rm H, \,1 cie + 2 cf}$ to an area coincident with the main filaments using Chandra spectra, 
then we find that $\dot M_{\rm H, \,1 cie + 2 cf}$ drops by a factor of 3 and there is agreement with the particle heating model. 
Inspection of other clusters with $\dot M_{\rm neb}<10 \Msunpyr$ shows that they also have soft emission more extended than the nebula. 
Note that \citet{2018arXiv180309765H} find weak optical [NII] line emission outside the filamentary nebula in the Centaurus cluster that, 
if common in other clusters, implies that the nebula emission is yet more extended than assumed above.

The nature of the extended 0.7\,keV gas component in the Centaurus cluster and other clusters on the left hand side of Fig. \ref{fig:mass_compare_1} is unclear. 
The low value of $\dot M_{\rm H, \,1 cie + 2 cf}$ shows that it is not rapidly radiatively cooling. 
We compare its luminosity $L_{\rm C, 2\, cie}$ (Table \ref{table:2cie}) with $L_{\rm neb}$ in Fig. \ref{fig:mass_comp_2}.

For completeness we now also consider the effect of different metallicity, 
in the light of the inner abundance gradient seen in the Centaurus (\citealt{2013MNRAS.433.3290P}; \citealt{2018MNRAS.481.4472L}) and other clusters (\citealt{2015MNRAS.447..417P}). 
This takes the form of a pronounced drop in Fe and other abundances within the innermost 10 kpc. 
In Centaurus the Fe abundance is about twice the Solar value at 15--20 kpc and drops to below 0.4 Solar within 5 kpc.
\citet{2013MNRAS.433.3290P} hypothesise that this is due to most metals in stellar mass-loss being in the form of grains which remain trapped in cold clouds. 
They are then transported out of the cluster centre by the bubbling feedback process and dumped at 10-20 kpc where they mix into the ICM. 
This means that the region where the ICM temperature is lowest and where a cooling flow might otherwise be expected has a low metallicity. 
A cooling flow would then produce only weak lines and any detection or limit would rely more on the continuum shape. 
As examples we have fitted the RGS spectra of the Centaurus cluster and A3581 with the two-stage cooling flow model setting $Z=0.05Z\sun$ for the second stage (cooling from 0.7 to 0.01\,keV). 
The rates for that stage are 15 and $24\,\Msunpyr$, respectively. 
We are not claiming that these are solutions for any continuous cooling flow as we have no reason to suspect that the abundance could drop as the temperature passed below 0.7\,keV. 
But the possibility remains that some cooling may occur within the coolest central gas, which may be in the process of being dragged out from the centre (\citealt{2013MNRAS.433.3290P}).

It is possible that intrinsic absorption could reduce the level of cooling of the lowest temperature components. 
The molecular nebula is a potential source of such obscuring gas. 
We have not included intrinsic absorption in this work and refer the reader to Fig. 18 in \citet{2008MNRAS.385.1186S} for the effect it has on the data from the Centaurus cluster. 
Results for other clusters in our sample will be relatively similar.

In summary, the current spectra neither support nor rule out the idea that the H$\alpha$/CO filaments are powered by interpenetration by the surrounding gas. 
It remains possible that if it does occur, then it can be from gas at around 0.7\,keV in most of the clusters studied but requires hotter gas at $T_{\rm Max, \,1 cie + 2 cf}$ for the most luminous objects.

\subsection{Star formation rates}
% \label{sec:sfr}

\begin{figure*} \includegraphics[width=\linewidth,
angle=0]{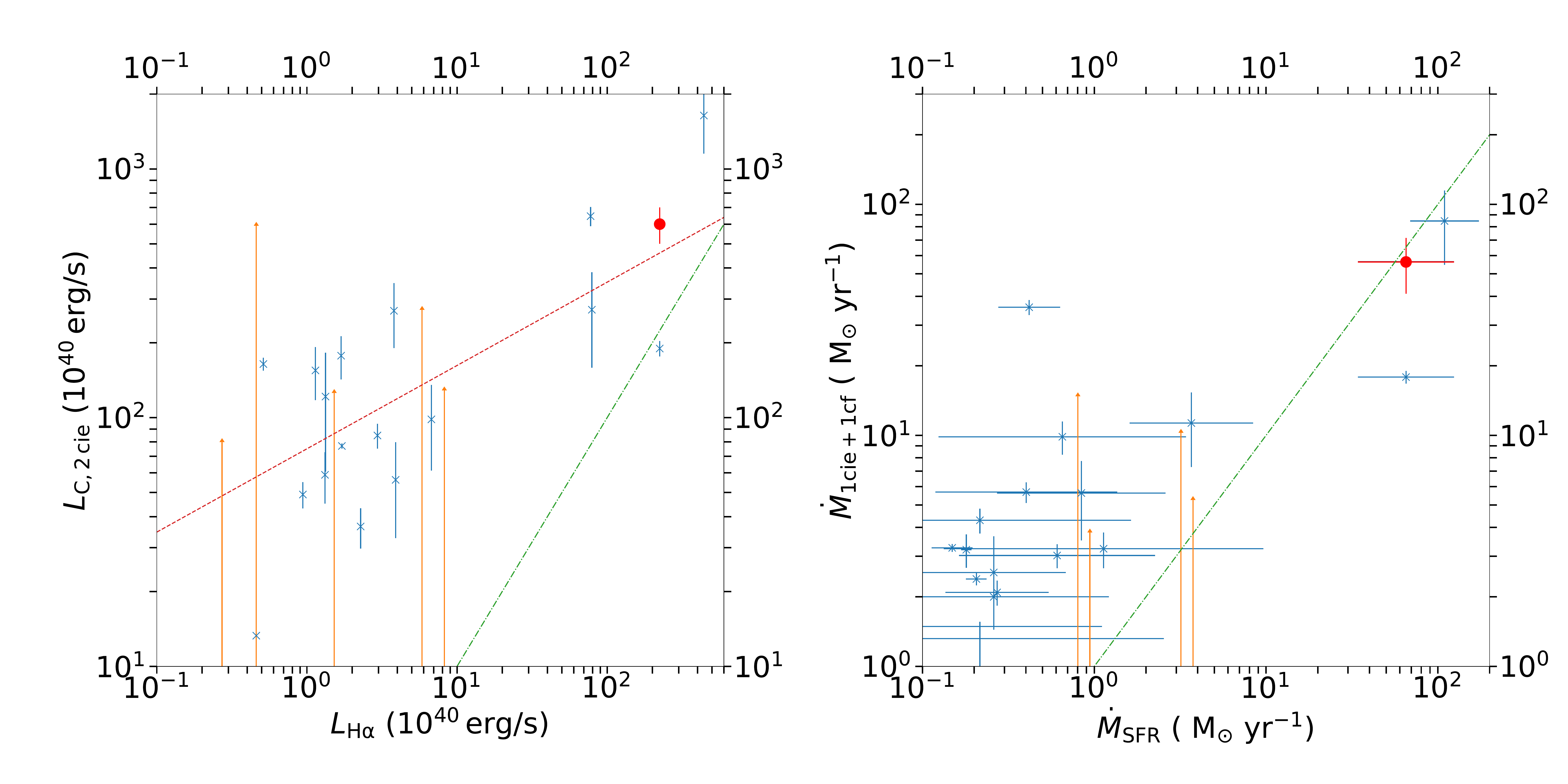}
  \caption{Left: The luminosity of the cooler \textit{cie} component in the
2 \textit{cie} model against the H$\alpha$ luminosity. The red
dash line represents the best linear fit to the data in the log space.
Right: Cooling rates against observed star formation rates (\citealt{2018ApJ...858...45M} references therein).  
The red circles are the Perseus cluster with 99$\%$ PSF spectrum in both panels.}
  \label{fig:mass_comp_2}
\end{figure*}

Another aspect we can investigate is whether there is a link between cooling rates and the observed star formation rates.
Assuming soft X-ray cooling flows lose their energy to the filaments and the cooled gas is consumed directly in star formation, 
their difference gives us hints on the rate of change in molecular mass. 
We compare these parameters in the right panel in Fig. \ref{fig:mass_comp_2}.
Most clusters and groups have a star formation rate lower than 5 $\Msunpyr$, 
which is 5 to more than 50 times lower than the measured cooling rate. 
Only for the more massive clusters Perseus and A1835 with strong star formation activity, 
the cooling rate close matches the star formation rate and the formation efficiency is around 80 \%. 
This efficiency is higher than the minimum star formation efficiency predicted by \citet{2018ApJ...858...45M} using the `simple' cooling rates. 
We find a weak trend of increasing in star formation efficiency with the cooling rate in agreement with \citet{2018ApJ...858...45M}.
Since H$\alpha$ filaments are not necessarily aligned with star formation regions (e.g. the Perseus cluster, \citealt{2010MNRAS.405..115C,2014MNRAS.444..336C}), 
the connection between cooling flows and star formation can be very complicated in massive clusters.
The measured cooling rates and upper limits are above the line of unity, and therefore the cooling flows give a net increase in their molecular mass,
which accumulates at a level of a few or a few tens of $\Msunpyr$. 
Assuming the mass of clusters is of the order of $10^{9}-10^{10}\rm M_{\odot}$ (e.g., $4\times 10^{10}\rm M_{\odot}$ in the Perseus cluster,
\citealt{2006A&A...454..437S}; $9\times 10^{8}\rm M_{\odot}$ in A262, \citealt{2001MNRAS.328..762E}), 
a constant mass accumulation rate means that the age of molecular gas is at around a few $10^{8}\rm yr$ to $10^{9}\rm yr$, 
which is comparable to the age of clusters.
Therefore, it is possible that no significant molecular gas content was present at sufficiently early epoch. 

\section{Conclusions}
\label{sec:conclusion} 
In this paper, we analysed the RGS spectra of the core of 45 clusters and groups of galaxies, and searched for cool and cooling gas. 
The continuum of the spectra are modelled by a collisional ionisation equilibrium (\textit{cie}) component, which has a typical bulk temperature of 1-4 keV. 
Since {Fe\,\scriptsize{XVII}} emission is observed, there is either a cooling flow or a cooler gas component. 
If cooling flows were taking place in these objects, we can measure the cooling rates with both the one-stage and two-stage cooling flow models. 
Alternatively, we search for cooler gas with the 2 \textit{cie} model. 
The results are as follows.

\begin{itemize}
 \item In the one-stage cooling flow model, all but one (AWM7) have $\dot M/\dot M_{\rm simple, 3 Gyr}$ less than 0.4 and 19 out of 22 clusters have the ratio less than 0.2. 
 \item In the two-stage cooling flow model, we measured the cooling rates between the bulk temperature and 0.7 keV. 
 We find that all clusters but one (A262) have cooling rates less than 70$\%$ of $\dot M_{\rm simple, 3 Gyr}$, and 17 out of 22 clusters have cooling rates less than 30$\%$ of $\dot M_{\rm simple, 3 Gyr}$.
 \item The residual cooling rates below 0.7 keV are less than 30$\%$ of $\dot M_{\rm simple, 3 Gyr}$ in all clusters except AWM7, and only 10$\%$ in 15/22 clusters. 
 Therefore we find no strong evidence that clusters are rapidly radiatively cooling below 0.7 keV. 
 which suggest that cooling flows appear to stop cooling at around 0.7 keV. 
 \item The 2 \textit{cie} model gives the cooler temperature between 0.5-0.9 keV in most clusters with the mean temperature of 0.78$\pm$0.13 keV for those higher than 0.4 keV. 
 \item In 9 out of 22 clusters, we have statistical evidence that the two-stage model provides a better spectral fit than the one-stage model ($\Delta$C-stat$\geq 10$ for 1 degree of freedom). 
 For most clusters, we cannot determine whether the 2 \textit{cie} model or the cooling flow models provide a significantly better spectral fit. 
\end{itemize}

Since the soft X-ray emission happens to be spatially associated with H$\alpha$ nebulosity, we investigated the relation between the cooling rates above 0.7 keV and the total optical-UV luminosities. 
We find that the detected cooling rates have enough energy to power the total optical-UV luminosities for low luminosity objects, where the soft X-ray region is more extended than the H$\alpha$ nebula. 
For the 4 high luminosity objects, we observe the opposite situation where the cooling rates are not sufficient. 
This suggests that if the X-ray cooling gas were powering the nebulae, it requires an inflow at a higher temperature. 
Finally, we find the cooling rates above 0.7 keV are 5 to 50 times higher than the observed star formation rates, 
which suggests that it is possible that the mass of the molecular gas reservoir is gradually increasing in most objects.

\vspace{-0.5cm}

%\clearpage
\section*{Acknowledgements}

This work is based on observations obtained with XMM-\textit{Newton},
an ESA science mission funded by ESA Member States and USA (NASA).  CP
and ACF acknowledge support from the European Research Council through
Advanced Grant on Feedback 340442.

\vspace{-0.5cm}

\bibliographystyle{mn2e}
\bibliography{Haonan_cluster_paper} %----> Haonan_cluster_paper.bib

\begin{thebibliography}{}

\bibitem[\protect\citeauthoryear{{Allen}, {Fabian}, {Johnstone}, {Arnaud} \&
  {Nulsen}}{{Allen} et~al.}{2001}]{2001MNRAS.322..589A}
{Allen} S.~W.,  {Fabian} A.~C.,  {Johnstone} R.~M.,  {Arnaud} K.~A.,
  {Nulsen} P.~E.~J.,  2001, \mnras, 322, 589

\bibitem[\protect\citeauthoryear{{Bambic}, {Pinto}, {Fabian}, {Sanders} \&
  {Reynolds}}{{Bambic} et~al.}{2018}]{2018MNRAS.478L..44B}
{Bambic} C.~J.,  {Pinto} C.,  {Fabian} A.~C.,  {Sanders} J.,    {Reynolds}
  C.~S.,  2018, \mnras, 478, L44

\bibitem[\protect\citeauthoryear{{Bauer}, {Fabian}, {Sanders}, {Allen} \&
  {Johnstone}}{{Bauer} et~al.}{2005}]{2005MNRAS.359.1481B}
{Bauer} F.~E.,  {Fabian} A.~C.,  {Sanders} J.~S.,  {Allen} S.~W.,
  {Johnstone} R.~M.,  2005, \mnras, 359, 1481

\bibitem[\protect\citeauthoryear{{Bregman}, {Miller}, {Athey} \&
  {Irwin}}{{Bregman} et~al.}{2005}]{2005ApJ...635.1031B}
{Bregman} J.~N.,  {Miller} E.~D.,  {Athey} A.~E.,    {Irwin} J.~A.,  2005,
  \apj, 635, 1031

\bibitem[\protect\citeauthoryear{{Canning}, {Fabian}, {Johnstone}, {Sanders},
  {Conselice}, {Crawford}, {Gallagher} \& {Zweibel}}{{Canning}
  et~al.}{2010}]{2010MNRAS.405..115C}
{Canning} R.~E.~A.,  {Fabian} A.~C.,  {Johnstone} R.~M.,  {Sanders} J.~S.,
  {Conselice} C.~J.,  {Crawford} C.~S.,  {Gallagher} J.~S.,    {Zweibel} E.,
  2010, \mnras, 405, 115

\bibitem[\protect\citeauthoryear{{Canning}, {Ryon}, {Gallagher}, {Kotulla},
  {O'Connell}, {Fabian}, {Johnstone}, {Conselice}, {Hicks}, {Rosario} \&
  {Wyse}}{{Canning} et~al.}{2014}]{2014MNRAS.444..336C}
{Canning} R.~E.~A.,  {Ryon} J.~E.,  {Gallagher} J.~S.,  {Kotulla} R.,
  {O'Connell} R.~W.,  {Fabian} A.~C.,  {Johnstone} R.~M.,  {Conselice} C.~J.,
  {Hicks} A.,  {Rosario} D.,    {Wyse} R.~F.~G.,  2014, \mnras, 444, 336

\bibitem[\protect\citeauthoryear{{Cavagnolo}, {Donahue}, {Voit} \&
  {Sun}}{{Cavagnolo} et~al.}{2008}]{2008ApJ...683L.107C}
{Cavagnolo} K.~W.,  {Donahue} M.,  {Voit} G.~M.,    {Sun} M.,  2008, \apjl,
  683, L107

\bibitem[\protect\citeauthoryear{{Cavagnolo}, {Donahue}, {Voit} \&
  {Sun}}{{Cavagnolo} et~al.}{2009}]{2009ApJS..182...12C}
{Cavagnolo} K.~W.,  {Donahue} M.,  {Voit} G.~M.,    {Sun} M.,  2009, \apjs,
  182, 12

\bibitem[\protect\citeauthoryear{{Chen}, {Reiprich}, {B{\"o}hringer}, {Ikebe}
  \& {Zhang}}{{Chen} et~al.}{2007}]{2007A&A...466..805C}
{Chen} Y.,  {Reiprich} T.~H.,  {B{\"o}hringer} H.,  {Ikebe} Y.,    {Zhang}
  Y.-Y.,  2007, \aap, 466, 805

\bibitem[\protect\citeauthoryear{{Churazov}, {Forman}, {Jones} \&
  {B{\"o}hringer}}{{Churazov} et~al.}{2003}]{2003ApJ...590..225C}
{Churazov} E.,  {Forman} W.,  {Jones} C.,    {B{\"o}hringer} H.,  2003, ApJ,
  590, 225

\bibitem[\protect\citeauthoryear{{Conselice}, {Gallagher} III \&
  {Wyse}}{{Conselice} et~al.}{2001}]{2001AJ....122.2281C}
{Conselice} C.~J.,  {Gallagher} III J.~S.,    {Wyse} R.~F.~G.,  2001, \aj, 122,
  2281

\bibitem[\protect\citeauthoryear{{Crawford}, {Allen}, {Ebeling}, {Edge} \&
  {Fabian}}{{Crawford} et~al.}{1999}]{1999MNRAS.306..857C}
{Crawford} C.~S.,  {Allen} S.~W.,  {Ebeling} H.,  {Edge} A.~C.,    {Fabian}
  A.~C.,  1999, \mnras, 306, 857

\bibitem[\protect\citeauthoryear{{Crawford}, {Hatch}, {Fabian} \&
  {Sanders}}{{Crawford} et~al.}{2005}]{2005MNRAS.363..216C}
{Crawford} C.~S.,  {Hatch} N.~A.,  {Fabian} A.~C.,    {Sanders} J.~S.,  2005,
  \mnras, 363, 216

\bibitem[\protect\citeauthoryear{{Crawford}, {Sanders} \& {Fabian}}{{Crawford}
  et~al.}{2005}]{2005MNRAS.361...17C}
{Crawford} C.~S.,  {Sanders} J.~S.,    {Fabian} A.~C.,  2005, \mnras, 361, 17

\bibitem[\protect\citeauthoryear{{David}, {Nulsen}, {McNamara}, {Forman},
  {Jones}, {Ponman}, {Robertson} \& {Wise}}{{David}
  et~al.}{2001}]{2001ApJ...557..546D}
{David} L.~P.,  {Nulsen} P.~E.~J.,  {McNamara} B.~R.,  {Forman} W.,  {Jones}
  C.,  {Ponman} T.,  {Robertson} B.,    {Wise} M.,  2001, \apj, 557, 546

\bibitem[\protect\citeauthoryear{{de Plaa}, {Kaastra}, {Werner}, {Pinto} \&
  {Kosec}}{{de Plaa} et~al.}{2017}]{2017A&A...607A..98D}
{de Plaa} J.,  {Kaastra} J.~S.,  {Werner} N.,  {Pinto} C.,    {Kosec} P. e.~a.,
   2017, A\&A, 607, A98

\bibitem[\protect\citeauthoryear{{Donahue}, {Sun}, {O'Dea}, {Voit} \&
  {Cavagnolo}}{{Donahue} et~al.}{2007}]{2007AJ....134...14D}
{Donahue} M.,  {Sun} M.,  {O'Dea} C.~P.,  {Voit} G.~M.,    {Cavagnolo} K.~W.,
  2007, \aj, 134, 14

\bibitem[\protect\citeauthoryear{{Donato}, {Sambruna} \& {Gliozzi}}{{Donato}
  et~al.}{2004}]{2004ApJ...617..915D}
{Donato} D.,  {Sambruna} R.~M.,    {Gliozzi} M.,  2004, ApJ, 617, 915

\bibitem[\protect\citeauthoryear{{Edge}}{{Edge}}{2001}]{2001MNRAS.328..762E}
{Edge} A.~C.,  2001, \mnras, 328, 762

\bibitem[\protect\citeauthoryear{{Fabian}}{{Fabian}}{1994}]{1994ARA&A..32..277F}
{Fabian} A.~C.,  1994, \araa, 32, 277

\bibitem[\protect\citeauthoryear{{Fabian}}{{Fabian}}{2012}]{2012ARA&A..50..455F}
{Fabian} A.~C.,  2012, \araa, 50, 455

\bibitem[\protect\citeauthoryear{{Fabian}, {Allen}, {Crawford}, {Johnstone},
  {Morris}, {Sanders} \& {Schmidt}}{{Fabian}
  et~al.}{2002}]{2002MNRAS.332L..50F}
{Fabian} A.~C.,  {Allen} S.~W.,  {Crawford} C.~S.,  {Johnstone} R.~M.,
  {Morris} R.~G.,  {Sanders} J.~S.,    {Schmidt} R.~W.,  2002, \mnras, 332, L50

\bibitem[\protect\citeauthoryear{{Fabian}, {Arnaud}, {Nulsen}, {Watson},
  {Stewart}, {McHardy}, {Smith}, {Cooke}, {Elvis} \& {Mushotzky}}{{Fabian}
  et~al.}{1985}]{1985MNRAS.216..923F}
{Fabian} A.~C.,  {Arnaud} K.~A.,  {Nulsen} P.~E.~J.,  {Watson} M.~G.,
  {Stewart} G.~C.,  {McHardy} I.,  {Smith} A.,  {Cooke} B.,  {Elvis} M.,
  {Mushotzky} R.~F.,  1985, \mnras, 216, 923

\bibitem[\protect\citeauthoryear{{Fabian}, {Johnstone}, {Sanders}, {Conselice},
  {Crawford}, {Gallagher} III \& {Zweibel}}{{Fabian}
  et~al.}{2008}]{2008Natur.454..968F}
{Fabian} A.~C.,  {Johnstone} R.~M.,  {Sanders} J.~S.,  {Conselice} C.~J.,
  {Crawford} C.~S.,  {Gallagher} III J.~S.,    {Zweibel} E.,  2008, Nature,
  454, 968

\bibitem[\protect\citeauthoryear{{Fabian}, {Reynolds}, {Taylor} \&
  {Dunn}}{{Fabian} et~al.}{2005}]{2005MNRAS.363..891F}
{Fabian} A.~C.,  {Reynolds} C.~S.,  {Taylor} G.~B.,    {Dunn} R.~J.~H.,  2005,
  \mnras, 363, 891

\bibitem[\protect\citeauthoryear{{Fabian}, {Sanders}, {Crawford}, {Conselice},
  {Gallagher} \& {Wyse}}{{Fabian} et~al.}{2003}]{2003MNRAS.344L..48F}
{Fabian} A.~C.,  {Sanders} J.~S.,  {Crawford} C.~S.,  {Conselice} C.~J.,
  {Gallagher} J.~S.,    {Wyse} R.~F.~G.,  2003, \mnras, 344, L48

\bibitem[\protect\citeauthoryear{{Fabian}, {Sanders}, {Ettori}, {Taylor},
  {Allen}, {Crawford}, {Iwasawa} \& {Johnstone}}{{Fabian}
  et~al.}{2001}]{2001MNRAS.321L..33F}
{Fabian} A.~C.,  {Sanders} J.~S.,  {Ettori} S.,  {Taylor} G.~B.,  {Allen}
  S.~W.,  {Crawford} C.~S.,  {Iwasawa} K.,    {Johnstone} R.~M.,  2001, \mnras,
  321, L33

\bibitem[\protect\citeauthoryear{{Fabian}, {Sanders}, {Taylor}, {Allen},
  {Crawford}, {Johnstone} \& {Iwasawa}}{{Fabian}
  et~al.}{2006}]{2006MNRAS.366..417F}
{Fabian} A.~C.,  {Sanders} J.~S.,  {Taylor} G.~B.,  {Allen} S.~W.,  {Crawford}
  C.~S.,  {Johnstone} R.~M.,    {Iwasawa} K.,  2006, \mnras, 366, 417

\bibitem[\protect\citeauthoryear{{Fabian}, {Sanders}, {Williams}, {Lazarian},
  {Ferland} \& {Johnstone}}{{Fabian} et~al.}{2011}]{2011MNRAS.417..172F}
{Fabian} A.~C.,  {Sanders} J.~S.,  {Williams} R.~J.~R.,  {Lazarian} A.,
  {Ferland} G.~J.,    {Johnstone} R.~M.,  2011, \mnras, 417, 172

\bibitem[\protect\citeauthoryear{{Fabian}, {Walker}, {Russell}, {Pinto},
  {Canning}, {Salome}, {Sanders}, {Taylor}, {Zweibel}, {Conselice}, {Combes},
  {Crawford}, {Ferland}, {Gallagher} III, {Hatch}, {Johnstone} \&
  {Reynolds}}{{Fabian} et~al.}{2016}]{2016MNRAS.461..922F}
{Fabian} A.~C.,  {Walker} S.~A.,  {Russell} H.~R.,  {Pinto} C.,  {Canning}
  R.~E.~A.,  {Salome} P.,  {Sanders} J.~S.,  {Taylor} G.~B.,  {Zweibel} E.~G.,
  {Conselice} C.~J.,  {Combes} F.,  {Crawford} C.~S.,  {Ferland} G.~J.,
  {Gallagher} III J.~S.,  {Hatch} N.~A.,  {Johnstone} R.~M.,    {Reynolds}
  C.~S.,  2016, \mnras, 461, 922

\bibitem[\protect\citeauthoryear{{Fabian}, {Walker}, {Russell}, {Pinto},
  {Sanders} \& {Reynolds}}{{Fabian} et~al.}{2017}]{2017MNRAS.464L...1F}
{Fabian} A.~C.,  {Walker} S.~A.,  {Russell} H.~R.,  {Pinto} C.,  {Sanders}
  J.~S.,    {Reynolds} C.~S.,  2017, \mnras, 464, L1

\bibitem[\protect\citeauthoryear{{Ferland}, {Fabian}, {Hatch}, {Johnstone},
  {Porter}, {van Hoof} \& {Williams}}{{Ferland}
  et~al.}{2009}]{2009MNRAS.392.1475F}
{Ferland} G.~J.,  {Fabian} A.~C.,  {Hatch} N.~A.,  {Johnstone} R.~M.,  {Porter}
  R.~L.,  {van Hoof} P.~A.~M.,    {Williams} R.~J.~R.,  2009, \mnras, 392, 1475

\bibitem[\protect\citeauthoryear{{Hamer}, {Edge}, {Swinbank}, {Wilman},
  {Combes}, {Salom{\'e}}, {Fabian}, {Crawford}, {Russell}, {Hlavacek-Larrondo},
  {McNamara} \& {Bremer}}{{Hamer} et~al.}{2016}]{2016MNRAS.460.1758H}
{Hamer} S.~L.,  {Edge} A.~C.,  {Swinbank} A.~M.,  {Wilman} R.~J.,  {Combes} F.,
   {Salom{\'e}} P.,  {Fabian} A.~C.,  {Crawford} C.~S.,  {Russell} H.~R.,
  {Hlavacek-Larrondo} J.,  {McNamara} B.~R.,    {Bremer} M.~N.,  2016, \mnras,
  460, 1758

\bibitem[\protect\citeauthoryear{{Hamer}, {Fabian}, {Russell}, {Salom{\'e}},
  {Combes}, {Olivares}, {Polles}, {Edge} \& {Beckmann}}{{Hamer}
  et~al.}{2018}]{2018arXiv180309765H}
{Hamer} S.~L.,  {Fabian} A.~C.,  {Russell} H.~R.,  {Salom{\'e}} P.,  {Combes}
  F.,  {Olivares} V.,  {Polles} F.~L.,  {Edge} A.~C.,    {Beckmann} R.~S.,
  2018, arXiv e-prints

\bibitem[\protect\citeauthoryear{{Heckman}, {Baum}, {van Breugel} \&
  {McCarthy}}{{Heckman} et~al.}{1989}]{1989ApJ...338...48H}
{Heckman} T.~M.,  {Baum} S.~A.,  {van Breugel} W.~J.~M.,    {McCarthy} P.,
  1989, \apj, 338, 48

\bibitem[\protect\citeauthoryear{{Hudson}, {Mittal}, {Reiprich}, {Nulsen},
  {Andernach} \& {Sarazin}}{{Hudson} et~al.}{2010}]{2010A&A...513A..37H}
{Hudson} D.~S.,  {Mittal} R.,  {Reiprich} T.~H.,  {Nulsen} P.~E.~J.,
  {Andernach} H.,    {Sarazin} C.~L.,  2010, \aap, 513, A37

\bibitem[\protect\citeauthoryear{{Jaffe}, {Bremer} \& {Baker}}{{Jaffe}
  et~al.}{2005}]{2005MNRAS.360..748J}
{Jaffe} W.,  {Bremer} M.~N.,    {Baker} K.,  2005, \mnras, 360, 748

\bibitem[\protect\citeauthoryear{{Johnstone}, {Fabian} \& {Nulsen}}{{Johnstone}
  et~al.}{1987}]{1987MNRAS.224...75J}
{Johnstone} R.~M.,  {Fabian} A.~C.,    {Nulsen} P.~E.~J.,  1987, \mnras, 224,
  75

\bibitem[\protect\citeauthoryear{{Kaastra}, {Ferrigno}, {Tamura}, {Paerels},
  {Peterson} \& {Mittaz}}{{Kaastra} et~al.}{2001}]{2001A&A...365L..99K}
{Kaastra} J.~S.,  {Ferrigno} C.,  {Tamura} T.,  {Paerels} F.~B.~S.,  {Peterson}
  J.~R.,    {Mittaz} J.~P.~D.,  2001, \aap, 365, L99

\bibitem[\protect\citeauthoryear{{Kaiser}}{{Kaiser}}{1991}]{1991ApJ...383..104K}
{Kaiser} N.,  1991, \apj, 383, 104

\bibitem[\protect\citeauthoryear{{Kalberla}, {Burton}, {Hartmann}, {Arnal},
  {Bajaja}, {Morras} \& {P{\"o}ppel}}{{Kalberla}
  et~al.}{2005}]{2005A&A...440..775K}
{Kalberla} P.~M.~W.,  {Burton} W.~B.,  {Hartmann} D.,  {Arnal} E.~M.,  {Bajaja}
  E.,  {Morras} R.,    {P{\"o}ppel} W.~G.~L.,  2005, \aap, 440, 775

\bibitem[\protect\citeauthoryear{{Lakhchaura}, {Werner}, {Sun}, {Canning},
  {Gaspari}, {Allen}, {Connor}, {Donahue} \& {Sarazin}}{{Lakhchaura}
  et~al.}{2018}]{2018MNRAS.481.4472L}
{Lakhchaura} K.,  {Werner} N.,  {Sun} M.,  {Canning} R.~E.~A.,  {Gaspari} M.,
  {Allen} S.~W.,  {Connor} T.,  {Donahue} M.,    {Sarazin} C.,  2018, \mnras,
  481, 4472

\bibitem[\protect\citeauthoryear{{Lodders} \& {Palme}}{{Lodders} \&
  {Palme}}{2009}]{2009M&PSA..72.5154L}
{Lodders} K.,  {Palme} H.,  2009, Meteoritics and Planetary Science Supplement,
  72, 5154

\bibitem[\protect\citeauthoryear{{McDonald}, {Gaspari}, {McNamara} \&
  {Tremblay}}{{McDonald} et~al.}{2018}]{2018ApJ...858...45M}
{McDonald} M.,  {Gaspari} M.,  {McNamara} B.~R.,    {Tremblay} G.~R.,  2018,
  \apj, 858, 45

\bibitem[\protect\citeauthoryear{{McNamara} \& {Nulsen}}{{McNamara} \&
  {Nulsen}}{2007}]{2007ARA&A..45..117M}
{McNamara} B.~R.,  {Nulsen} P.~E.~J.,  2007, \araa, 45, 117

\bibitem[\protect\citeauthoryear{{McNamara} \& {Nulsen}}{{McNamara} \&
  {Nulsen}}{2012}]{2012NJPh...14e5023M}
{McNamara} B.~R.,  {Nulsen} P.~E.~J.,  2012, New Journal of Physics, 14, 055023

\bibitem[\protect\citeauthoryear{{Mushotzky} \& {Szymkowiak}}{{Mushotzky} \&
  {Szymkowiak}}{1988}]{Mushotzky}
{Mushotzky} R.~F.,  {Szymkowiak} A.~E.,  1988, {Cooling Flows in Clusters and
  Galaxies (Nato ASI Series C:)}.
Kluwer Academic Publishers

\bibitem[\protect\citeauthoryear{{Nulsen}, {Johnstone} \& {Fabian}}{{Nulsen}
  et~al.}{1987}]{1987PASAu...7..132N}
{Nulsen} P.~E.~J.,  {Johnstone} R.~M.,    {Fabian} A.~C.,  1987, Proceedings of
  the Astronomical Society of Australia, 7, 132

\bibitem[\protect\citeauthoryear{{O'Dea}, {Baum}, {Privon}, {Noel-Storr},
  {Quillen}, {Zufelt}, {Park}, {Edge}, {Russell}, {Fabian}, {Donahue},
  {Sarazin}, {McNamara}, {Bregman} \& {Egami}}{{O'Dea}
  et~al.}{2008}]{2008ApJ...681.1035O}
{O'Dea} C.~P.,  {Baum} S.~A.,  {Privon} G.,  {Noel-Storr} J.,  {Quillen} A.~C.,
   {Zufelt} N.,  {Park} J.,  {Edge} A.,  {Russell} H.,  {Fabian} A.~C.,
  {Donahue} M.,  {Sarazin} C.~L.,  {McNamara} B.,  {Bregman} J.~N.,    {Egami}
  E.,  2008, \apj, 681, 1035

\bibitem[\protect\citeauthoryear{{Panagoulia}, {Fabian} \&
  {Sanders}}{{Panagoulia} et~al.}{2013}]{2013MNRAS.433.3290P}
{Panagoulia} E.~K.,  {Fabian} A.~C.,    {Sanders} J.~S.,  2013, \mnras, 433,
  3290

\bibitem[\protect\citeauthoryear{{Panagoulia}, {Fabian} \&
  {Sanders}}{{Panagoulia} et~al.}{2014}]{2014MNRAS.438.2341P}
{Panagoulia} E.~K.,  {Fabian} A.~C.,    {Sanders} J.~S.,  2014, \mnras, 438,
  2341

\bibitem[\protect\citeauthoryear{{Panagoulia}, {Sanders} \&
  {Fabian}}{{Panagoulia} et~al.}{2015}]{2015MNRAS.447..417P}
{Panagoulia} E.~K.,  {Sanders} J.~S.,    {Fabian} A.~C.,  2015, \mnras, 447,
  417

\bibitem[\protect\citeauthoryear{{Peres}, {Fabian}, {Edge}, {Allen},
  {Johnstone} \& {White}}{{Peres} et~al.}{1998}]{1998MNRAS.298..416P}
{Peres} C.~B.,  {Fabian} A.~C.,  {Edge} A.~C.,  {Allen} S.~W.,  {Johnstone}
  R.~M.,    {White} D.~A.,  1998, \mnras, 298, 416

\bibitem[\protect\citeauthoryear{{Peterson}, {Kahn}, {Paerels}, {Kaastra},
  {Tamura}, {Bleeker}, {Ferrigno} \& {Jernigan}}{{Peterson}
  et~al.}{2003}]{2003ApJ...590..207P}
{Peterson} J.~R.,  {Kahn} S.~M.,  {Paerels} F.~B.~S.,  {Kaastra} J.~S.,
  {Tamura} T.,  {Bleeker} J.~A.~M.,  {Ferrigno} C.,    {Jernigan} J.~G.,  2003,
  \apj, 590, 207

\bibitem[\protect\citeauthoryear{{Peterson}, {Paerels}, {Kaastra}, {Arnaud},
  {Reiprich}, {Fabian}, {Mushotzky}, {Jernigan} \& {Sakelliou}}{{Peterson}
  et~al.}{2001}]{2001A&A...365L.104P}
{Peterson} J.~R.,  {Paerels} F.~B.~S.,  {Kaastra} J.~S.,  {Arnaud} M.,
  {Reiprich} T.~H.,  {Fabian} A.~C.,  {Mushotzky} R.~F.,  {Jernigan} J.~G.,
  {Sakelliou} I.,  2001, \aap, 365, L104

\bibitem[\protect\citeauthoryear{{Pinto}, {Bambic}, {Sanders}, {Fabian},
  {McDonald}, {Russell}, {Liu} \& {Reynolds}}{{Pinto}
  et~al.}{2018}]{2018MNRAS.480.4113P}
{Pinto} C.,  {Bambic} C.~J.,  {Sanders} J.~S.,  {Fabian} A.~C.,  {McDonald} M.,
   {Russell} H.~R.,  {Liu} H.,    {Reynolds} C.~S.,  2018, \mnras, 480, 4113

\bibitem[\protect\citeauthoryear{{Pinto}, {Fabian}, {Ogorzalek}, {Zhuravleva},
  {Werner}, {Sanders}, {Zhang}, {Gu}, {de Plaa}, {Ahoranta}, {Finoguenov},
  {Johnstone} \& {Canning}}{{Pinto} et~al.}{2016}]{2016MNRAS.461.2077P}
{Pinto} C.,  {Fabian} A.~C.,  {Ogorzalek} A.,  {Zhuravleva} I.,  {Werner} N.,
  {Sanders} J.,  {Zhang} Y.-Y.,  {Gu} L.,  {de Plaa} J.,  {Ahoranta} J.,
  {Finoguenov} A.,  {Johnstone} R.,    {Canning} R.~E.~A.,  2016, \mnras, 461,
  2077

\bibitem[\protect\citeauthoryear{{Pinto}, {Fabian}, {Werner}, {Kosec},
  {Ahoranta}, {de Plaa}, {Kaastra}, {Sanders}, {Zhang} \& {Finoguenov}}{{Pinto}
  et~al.}{2014}]{2014A&A...572L...8P}
{Pinto} C.,  {Fabian} A.~C.,  {Werner} N.,  {Kosec} P.,  {Ahoranta} J.,  {de
  Plaa} J.,  {Kaastra} J.~S.,  {Sanders} J.~S.,  {Zhang} Y.-Y.,    {Finoguenov}
  A.,  2014, \aap, 572, L8

\bibitem[\protect\citeauthoryear{{Pinto}, {Kaastra}, {Costantini} \& {de
  Vries}}{{Pinto} et~al.}{2013}]{2013A&A...551A..25P}
{Pinto} C.,  {Kaastra} J.~S.,  {Costantini} E.,    {de Vries} C.,  2013, \aap,
  551, A25

\bibitem[\protect\citeauthoryear{{Pinto}, {Sanders}, {Werner}, {de Plaa},
  {Fabian}, {Zhang}, {Kaastra}, {Finoguenov} \& {Ahoranta}}{{Pinto}
  et~al.}{2015}]{2015A&A...575A..38P}
{Pinto} C.,  {Sanders} J.~S.,  {Werner} N.,  {de Plaa} J.,  {Fabian} A.~C.,
  {Zhang} Y.-Y.,  {Kaastra} J.~S.,  {Finoguenov} A.,    {Ahoranta} J.,  2015,
  \aap, 575, A38

\bibitem[\protect\citeauthoryear{{Rafferty}, {McNamara} \& {Nulsen}}{{Rafferty}
  et~al.}{2008}]{2008ApJ...687..899R}
{Rafferty} D.~A.,  {McNamara} B.~R.,    {Nulsen} P.~E.~J.,  2008, \apj, 687,
  899

\bibitem[\protect\citeauthoryear{{Salom{\'e}} \& {Combes}}{{Salom{\'e}} \&
  {Combes}}{2003}]{2003A&A...412..657S}
{Salom{\'e}} P.,  {Combes} F.,  2003, \aap, 412, 657

\bibitem[\protect\citeauthoryear{{Salom{\'e}}, {Combes}, {Edge}, {Crawford},
  {Erlund}, {Fabian}, {Hatch}, {Johnstone}, {Sanders} \& {Wilman}}{{Salom{\'e}}
  et~al.}{2006}]{2006A&A...454..437S}
{Salom{\'e}} P.,  {Combes} F.,  {Edge} A.~C.,  {Crawford} C.,  {Erlund} M.,
  {Fabian} A.~C.,  {Hatch} N.~A.,  {Johnstone} R.~M.,  {Sanders} J.~S.,
  {Wilman} R.~J.,  2006, \aap, 454, 437

\bibitem[\protect\citeauthoryear{{Sanders} \& {Fabian}}{{Sanders} \&
  {Fabian}}{2011}]{2011MNRAS.412L..35S}
{Sanders} J.~S.,  {Fabian} A.~C.,  2011, \mnras, 412, L35

\bibitem[\protect\citeauthoryear{{Sanders} \& {Fabian}}{{Sanders} \&
  {Fabian}}{2012}]{2012MNRAS.421..726S}
{Sanders} J.~S.,  {Fabian} A.~C.,  2012, \mnras, 421, 726

\bibitem[\protect\citeauthoryear{{Sanders} \& {Fabian}}{{Sanders} \&
  {Fabian}}{2013}]{2013MNRAS.429.2727S}
{Sanders} J.~S.,  {Fabian} A.~C.,  2013, \mnras, 429, 2727

\bibitem[\protect\citeauthoryear{{Sanders}, {Fabian}, {Allen}, {Morris},
  {Graham} \& {Johnstone}}{{Sanders} et~al.}{2008}]{2008MNRAS.385.1186S}
{Sanders} J.~S.,  {Fabian} A.~C.,  {Allen} S.~W.,  {Morris} R.~G.,  {Graham}
  J.,    {Johnstone} R.~M.,  2008, \mnras, 385, 1186

\bibitem[\protect\citeauthoryear{{Sanders}, {Fabian}, {Frank}, {Peterson} \&
  {Russell}}{{Sanders} et~al.}{2010}]{2010MNRAS.402..127S}
{Sanders} J.~S.,  {Fabian} A.~C.,  {Frank} K.~A.,  {Peterson} J.~R.,
  {Russell} H.~R.,  2010, \mnras, 402, 127

\bibitem[\protect\citeauthoryear{{Sanders}, {Fabian} \& {Smith}}{{Sanders}
  et~al.}{2011}]{2011MNRAS.410.1797S}
{Sanders} J.~S.,  {Fabian} A.~C.,    {Smith} R.~K.,  2011, \mnras, 410, 1797

\bibitem[\protect\citeauthoryear{{Sanders}, {Fabian}, {Smith} \&
  {Peterson}}{{Sanders} et~al.}{2010}]{2010MNRAS.402L..11S}
{Sanders} J.~S.,  {Fabian} A.~C.,  {Smith} R.~K.,    {Peterson} J.~R.,  2010,
  \mnras, 402, L11

\bibitem[\protect\citeauthoryear{{Sanders}, {Fabian}, {Taylor}, {Russell},
  {Blundell}, {Canning}, {Hlavacek-Larrondo}, {Walker} \& {Grimes}}{{Sanders}
  et~al.}{2016}]{2016MNRAS.457...82S}
{Sanders} J.~S.,  {Fabian} A.~C.,  {Taylor} G.~B.,  {Russell} H.~R.,
  {Blundell} K.~M.,  {Canning} R.~E.~A.,  {Hlavacek-Larrondo} J.,  {Walker}
  S.~A.,    {Grimes} C.~K.,  2016, \mnras, 457, 82

\bibitem[\protect\citeauthoryear{{Snowden}, {Mushotzky}, {Kuntz} \&
  {Davis}}{{Snowden} et~al.}{2008}]{2008A&A...478..615S}
{Snowden} S.~L.,  {Mushotzky} R.~F.,  {Kuntz} K.~D.,    {Davis} D.~S.,  2008,
  \aap, 478, 615

\bibitem[\protect\citeauthoryear{{Stewart}, {Fabian}, {Jones} \&
  {Forman}}{{Stewart} et~al.}{1984}]{1984ApJ...285....1S}
{Stewart} G.~C.,  {Fabian} A.~C.,  {Jones} C.,    {Forman} W.,  1984, \apj,
  285, 1

\bibitem[\protect\citeauthoryear{{Tamura}, {Kaastra}, {Peterson}, {Paerels},
  {Mittaz}, {Trudolyubov}, {Stewart}, {Fabian}, {Mushotzky}, {Lumb} \&
  {Ikebe}}{{Tamura} et~al.}{2001}]{2001A&A...365L..87T}
{Tamura} T.,  {Kaastra} J.~S.,  {Peterson} J.~R.,  {Paerels} F.~B.~S.,
  {Mittaz} J.~P.~D.,  {Trudolyubov} S.~P.,  {Stewart} G.,  {Fabian} A.~C.,
  {Mushotzky} R.~F.,  {Lumb} D.~H.,    {Ikebe} Y.,  2001, \aap, 365, L87

\bibitem[\protect\citeauthoryear{{Voigt} \& {Fabian}}{{Voigt} \&
  {Fabian}}{2004}]{2004MNRAS.347.1130V}
{Voigt} L.~M.,  {Fabian} A.~C.,  2004, \mnras, 347, 1130

\bibitem[\protect\citeauthoryear{{Voit}, {Bryan}, {Balogh} \& {Bower}}{{Voit}
  et~al.}{2002}]{2002ApJ...576..601V}
{Voit} G.~M.,  {Bryan} G.~L.,  {Balogh} M.~L.,    {Bower} R.~G.,  2002, \apj,
  576, 601

\bibitem[\protect\citeauthoryear{{Walker}, {Kosec}, {Fabian} \&
  {Sanders}}{{Walker} et~al.}{2015}]{2015MNRAS.453.2480W}
{Walker} S.~A.,  {Kosec} P.,  {Fabian} A.~C.,    {Sanders} J.~S.,  2015,
  \mnras, 453, 2480

\bibitem[\protect\citeauthoryear{{Werner}, {Simionescu}, {Million}, {Allen},
  {Nulsen}, {von der Linden}, {Hansen}, {B{\"o}hringer}, {Churazov}, {Fabian},
  {Forman}, {Jones}, {Sanders} \& {Taylor}}{{Werner}
  et~al.}{2010}]{2010MNRAS.407.2063W}
{Werner} N.,  {Simionescu} A.,  {Million} E.~T.,  {Allen} S.~W.,  {Nulsen}
  P.~E.~J.,  {von der Linden} A.,  {Hansen} S.~M.,  {B{\"o}hringer} H.,
  {Churazov} E.,  {Fabian} A.~C.,  {Forman} W.~R.,  {Jones} C.,  {Sanders}
  J.~S.,    {Taylor} G.~B.,  2010, \mnras, 407, 2063

\bibitem[\protect\citeauthoryear{{White}, {Jones} \& {Forman}}{{White}
  et~al.}{1997}]{1997MNRAS.292..419W}
{White} D.~A.,  {Jones} C.,    {Forman} W.,  1997, \mnras, 292, 419

\bibitem[\protect\citeauthoryear{{Willingale}, {Starling}, {Beardmore},
  {Tanvir} \& {O'Brien}}{{Willingale} et~al.}{2013}]{2013MNRAS.431..394W}
{Willingale} R.,  {Starling} R.~L.~C.,  {Beardmore} A.~P.,  {Tanvir} N.~R.,
  {O'Brien} P.~T.,  2013, \mnras, 431, 394

\bibitem[\protect\citeauthoryear{{Wilman}, {Edge} \& {Swinbank}}{{Wilman}
  et~al.}{2006}]{2006MNRAS.371...93W}
{Wilman} R.~J.,  {Edge} A.~C.,    {Swinbank} A.~M.,  2006, \mnras, 371, 93

\bibitem[\protect\citeauthoryear{{Wright}}{{Wright}}{2006}]{2006PASP..118.1711W}
{Wright} E.~L.,  2006, \pasp, 118, 1711

\bibitem[\protect\citeauthoryear{{Wu}, {Fabian} \& {Nulsen}}{{Wu}
  et~al.}{2000}]{2000MNRAS.318..889W}
{Wu} K.~K.~S.,  {Fabian} A.~C.,    {Nulsen} P.~E.~J.,  2000, \mnras, 318, 889

\end{thebibliography}
\appendix
\section{}
\begin{table*}
\caption{Goodness of fits of different models. In each column, the values are C-statistic/degrees of freedom and their ratio.}  
 \vspace{-0.25cm}
\label{table:chi2}      % is used to refer this table in the text
\renewcommand{\arraystretch}{1.1}
% \begin{center}
%  \small\addtolength{\tabcolsep}{+2pt}
 
\scalebox{1.1}{%
\hspace*{-1cm}\begin{tabular}{c c c c c c c c c}     
\hline\hline            
Source                & 1 \textit{cie}  &  2 \textit{cie}  & 1 \textit{cie} + 1 \textit{cf}  & 1 \textit{cie} + 2 \textit{cf}   \\ \hline
2A0335+096            &  627/408  = 1.54  &  437/406  = 1.08  &  459/407  = 1.13 &  436/406  = 1.07 \\         
A85                   &  508/406  = 1.25  &  505/404  = 1.25  &  507/405  = 1.25 &  505/404  = 1.25 \\         
A133                  &  509/407  = 1.25  &  453/405  = 1.12  &  469/406  = 1.16 &  451/405  = 1.11 \\         
A262                  &  653/408  = 1.60  &  449/406  = 1.11  &  466/407  = 1.15 &  440/406  = 1.08 \\         
Perseus 90$\%$ PSF    &  940/408  = 2.30  &  723/406  = 1.78  &  758/407  = 1.86 &  727/406  = 1.79 \\         
Perseus 99$\%$ PSF    &  2246/408 = 5.50  &  1490/406 = 3.67  &  1597/407 = 3.92 &  1554/406 = 3.83 \\         
A496                  &  489/409  = 1.20  &  456/407  = 1.12  &  468/408  = 1.15 &  460/407  = 1.13 \\         
A1795                 &  432/408  = 1.06  &  427/406  = 1.05  &  431/407  = 1.06 &  430/406  = 1.06 \\         
A1835                 &  492/409  = 1.20  &  480/407  = 1.18  &  485/408  = 1.19 &  485/407  = 1.19 \\         
A1991                 &  434/406  = 1.07  &  394/404  = 0.98  &  400/405  = 0.99 &  393/404  = 0.97 \\         
A2029                 &  427/408  = 1.05  &  426/406  = 1.05  &  427/407  = 1.05 &  427/406  = 1.05 \\         
A2052                 &  530/409  = 1.29  &  442/407  = 1.09  &  474/408  = 1.16 &  442/407  = 1.09 \\         
A2199                 &  534/407  = 1.31  &  500/405  = 1.24  &  505/406  = 1.24 &  504/405  = 1.24 \\              
A2597                 &  502/409  = 1.23  &  491/407  = 1.21  &  495/408  = 1.21 &  492/407  = 1.21 \\                   
A2626                 &  431/408  = 1.06  &  429/406  = 1.06  &  430/407  = 1.06 &  429/406  = 1.06 \\                   
A3112                 &  527/408  = 1.29  &  517/406  = 1.27  &  520/407  = 1.28 &  518/406  = 1.28 \\                   
Centaurus             &  1952/407 = 4.80  &  579/403  = 1.44  &  885/406  = 2.18 &  694/405  = 1.71 \\                                
A3581                 &  561/406  = 1.38  &  502/404  = 1.24  &  495/405  = 1.22 &  494/404  = 1.22 \\         
A4038                 &  465/407  = 1.14  &  450/405  = 1.11  &  453/406  = 1.12 &  451/405  = 1.11 \\                     
A4059                 &  462/409  = 1.13  &  446/407  = 1.10  &  449/408  = 1.10 &  446/407  = 1.10 \\                     
AS1101                &  461/409  = 1.13  &  458/407  = 1.13  &  460/408  = 1.13 &  459/407  = 1.13 \\                        
AWM7                  &  542/407  = 1.33  &  479/405  = 1.18  &  486/406  = 1.20 &  486/405  = 1.20 \\                     
EXO0422-086           &  474/405  = 1.17  &  458/403  = 1.14  &  459/404  = 1.14 &  458/403  = 1.14 \\                     
Fornax                &  747/407  = 1.84  &  610/405  = 1.51  &  745/406  = 1.83 &  665/405  = 1.64 \\                     
Hydra A               &  394/409  = 0.96  &  384/407  = 0.94  &  384/408  = 0.94 &  384/407  = 0.94 \\                     
Virgo                 &  2988/1397= 2.14  &  2515/1395= 1.80  &  2633/1396= 1.89 &  2516/1395= 1.80 \\           
MKW3s                 &  501/408  = 1.23  &  489/406  = 1.21  &  496/407  = 1.22 &  492/406  = 1.21 \\                   
MKW4                  &  442/407  = 1.09  &  422/405  = 1.04  &  440/406  = 1.08 &  430/405  = 1.06 \\ \hline          
HCG62                 &  594/407  = 1.46  &  510/405  = 1.26  &  578/406  = 1.42 &                  \\             
NGC5044               &  696/406  = 1.71  &  669/404  = 1.66  &  696/405  = 1.72 &                  \\           
M49                   &  630/405  = 1.56  &                   &  619/404  = 1.53 &                  \\             
M86                   &  536/405  = 1.32  &                   &  532/404  = 1.32 &                  \\             
M89                   &  532/406  = 1.31  &                   &  530/405  = 1.31 &                  \\                    
NGC507                &  484/405  = 1.20  &                   &  440/404  = 1.09 &                  \\                      
NGC533                &  522/404  = 1.29  &                   &  522/403  = 1.30 &                  \\                      
NGC1316               &  775/406  = 1.91  &                   &  709/405  = 1.75 &                  \\                      
NGC1404               &  539/406  = 1.33  &                   &  537/405  = 1.33 &                  \\                      
NGC1550               &  541/407  = 1.33  &                   &  528/406  = 1.30 &                  \\                      
NGC3411               &  492/406  = 1.21  &                   &  492/405  = 1.21 &                  \\                          
NGC4261               &  591/406  = 1.46  &                   &  591/405  = 1.46 &                  \\                      
NGC4325               &  422/406  = 1.04  &                   &  422/405  = 1.04 &                  \\                      
NGC4374               &  620/404  = 1.53  &                   &  616/403  = 1.53 &                  \\                                  
NGC4636               &  839/409  = 2.05  &                   &  807/408  = 1.98 &                  \\                                      
NGC4649               &  671/406  = 1.65  &                   &  671/405  = 1.66 &                  \\                                       
NGC5813               &  836/407  = 2.05  &                   &  836/406  = 2.06 &                  \\                                          
NGC5846               &  800/407  = 1.97  &                   &  772/406  = 1.90 &                  \\                                          
                                                                                                                                                                                                                         
\hline                
\end{tabular}}
% \end{center}

\end{table*}

\begin{table*}
\caption{Relative abundances.}
\label{A2}

\hspace*{-0.5cm}\begin{tabular}{ccccccc}
\hline\hline
Object         & 1 \textit{cie} &&&2 \textit{cie} &&  \\\hline
               &O/Fe& Mg/Fe& Fe  &O/Fe&Mg/Fe& Fe  \\\hline
2A0335+096     &0.92& 0.71& 0.33&0.75&0.80&0.57 \\            
A85            &0.57& 0.59& 0.51&0.56&0.58&0.54 \\           
A133           &0.57& 0.63& 0.68&0.54&0.61&0.87 \\           
A262           &0.72& 0.50& 0.36&0.66&0.67&0.61 \\           
Perseus (90PSF)&1.23& 0.50& 0.20&0.98&0.64&0.32 \\           
Perseus (99PSF)&1.02& 0.48& 0.21&0.84&0.43&0.32 \\           
A496           &0.71& 0.81& 0.49&0.69&0.78&0.57 \\           
A1795          &1.02& 0.99& 0.35&0.80&0.98&0.38 \\           
A1835          &0.85& 1.05& 0.31&0.78&0.94&0.37 \\           
A1991          &0.78& 0.41& 0.49&0.66&0.44&0.76 \\           
A2029          &1.28& 0.46& 0.22&1.26&0.45&0.22 \\           
A2052          &0.71& 0.84& 0.44&0.64&0.83&0.63 \\           
A2199          &0.73& 0.90& 0.43&0.70&0.86&0.50 \\           
A2597          &0.90& 0.80& 0.37&0.83&0.77&0.42 \\           
A2626          &0.51& 0.40& 0.81&0.50&0.36&0.92 \\           
A3112          &0.64& 0.41& 0.52&0.62&0.41&0.56 \\           
A3526          &0.75& 0.24& 0.42&0.57&0.54&1.07 \\           
A3581          &0.86& 0.70& 0.38&0.80&0.76&0.47 \\           
A4038          &0.86& 0.37& 0.41&0.82&0.40&0.50 \\           
A4059          &0.49& 0.68& 0.63&0.48&0.66&0.70 \\           
AS1101         &0.65& 0.57& 0.40&0.61&0.57&0.41 \\           
AWM7           &1.08& 0.38& 0.26&0.83&0.48&0.46 \\           
EXO0422-086    &0.87& 0.85& 0.71&0.70&0.73&0.68 \\           
Fornax         &0.70& 1.46& 0.51&0.51&0.82&1.22 \\           
HYDRA          &0.85& 0.39& 0.30&0.81&0.42&0.30 \\           
M87            &0.80& 0.34& 0.42&0.71&0.46&0.58 \\           
MKW3s          &0.63& 0.49& 0.40&0.50&0.49&0.44 \\           
MKW4           &0.57& 0.74& 1.02&0.53&0.77&1.39 \\ \hline    
HCG62          &0.78& 1.94& 0.38&0.66&1.32&0.59 \\           
NGC5044        &0.87& 1.20& 0.47&0.82&1.03&0.55 \\           
M49            &0.79& 1.32& 0.61&/   &/   &/    \\           
M86            &1.29& 1.62& 0.28&/   &/   &/    \\    
M89            &1.77& 1.64& 0.16&/   &/   &/    \\                                                                                                 
NGC507         &0.67& 1.19& 0.66&/   &/   &/    \\                                                                                                 
NGC533         &0.80& 1.54& 0.76&/   &/   &/    \\                                                                                                 
NGC1316        &1.79& 2.28& 0.36&/   &/   &/    \\                                                                                                 
NGC1404        &1.07& 0.87& 0.38&/   &/   &/    \\                                                                                                 
NGC1550        &0.73& 0.56& 0.44&/   &/   &/    \\                                                                                                 
NGC3411        &0.54& 1.36& 1.04&/   &/   &/    \\                                                                                                 
NGC4261        &1.11& 2.15& 0.30&/   &/   &/    \\                                                                                                 
NGC4325        &0.61& 1.18& 0.62&/   &/   &/    \\                                                                                                 
NGC4374        &1.56& 1.71& 0.21&/   &/   &/    \\                                                                                                 
NGC4636        &1.11& 0.92& 0.32&/   &/   &/    \\                                                                                                 
NGC4649        &0.94& 0.97& 0.45&/   &/   &/    \\        	                                                                                      
NGC5813        &0.94& 0.91& 0.46&/   &/   &/    \\           		                                                                      
NGC5846        &1.23& 1.16& 0.45&/   &/   &/    \\                                                                                                 

\hline
\end{tabular}
\end{table*}

\begin{table*}
\caption{Relative abundances continued.}
\label{A3}

\hspace*{-0.5cm}\begin{tabular}{ccccccc}
\hline\hline
Object         & 1 \textit{cie} + 1 \textit{cf} &&&1 \textit{cie} + 2 \textit{cf} &&  \\\hline
               &O/Fe& Mg/Fe& Fe  &O/Fe&Mg/Fe& Fe  \\\hline
2A0335+096     &0.74&0.77&0.50&0.71&0.77&0.61 \\            
A85            &0.56&0.58&0.53&0.56&0.57&0.53 \\           
A133           &0.53&0.62&0.84&0.54&0.57&0.90 \\           
A262           &0.61&0.66&0.57&0.63&0.65&0.65 \\           
Perseus (90PSF)&0.91&0.55&0.33&0.92&0.62&0.30 \\           
Perseus (99PSF)&0.78&0.35&0.33&0.79&0.37&0.30 \\           
A496           &0.67&0.79&0.56&0.69&0.75&0.57 \\           
A1795          &0.98&0.95&0.37&0.95&0.96&0.37 \\           
A1835          &0.73&0.92&0.38&0.74&0.91&0.38 \\           
A1991          &0.64&0.45&0.68&0.64&0.42&0.79 \\           
A2029          &1.26&0.45&0.22&1.27&0.45&0.22 \\           
A2052          &0.64&0.85&0.55&0.64&0.81&0.64 \\           
A2199          &0.68&0.82&0.51&0.69&0.79&0.50 \\           
A2597          &0.83&0.78&0.41&0.85&0.75&0.42 \\           
A2626          &0.50&0.39&0.89&0.50&0.36&0.94 \\           
A3112          &0.62&0.40&0.55&0.61&0.41&0.55 \\           
A3526          &0.57&0.39&0.76&0.51&0.49&1.21 \\           
A3581          &0.70&0.75&0.48&0.69&0.77&0.51 \\           
A4038          &0.80&0.33&0.50&0.77&0.40&0.50 \\           
A4059          &0.47&0.67&0.70&0.48&0.64&0.71 \\           
AS1101         &0.65&0.56&0.40&0.64&0.56&0.40 \\           
AWM7           &0.79&0.39&0.45&0.81&0.40&0.46 \\           
EXO0422-086    &0.72&0.73&0.68&0.71&0.73&0.68 \\           
Fornax         &0.65&1.38&0.53&0.58&1.03&0.87 \\           
HYDRA          &0.82&0.42&0.30&0.82&0.42&0.30 \\           
M87            &0.66&0.38&0.54&0.70&0.45&0.58 \\           
MKW3s          &0.60&0.47&0.43&0.59&0.48&0.44 \\           
MKW4           &0.56&0.75&1.08&0.55&0.79&1.24 \\ \hline    
HCG62          &0.60&1.56&0.43&/   &/   &/    \\
NGC5044        &0.87&1.20&0.47&/   &/   &/    \\
M49            &0.67&1.14&0.68&/   &/   &/    \\
M86            &1.01&1.29&0.33&/   &/   &/    \\
M89            &1.33&1.65&0.18&/   &/   &/    \\
NGC507         &0.53&0.94&1.08&/   &/   &/    \\
NGC533         &0.78&1.51&0.77&/   &/   &/    \\
NGC1316        &0.87&0.96&0.81&/   &/   &/    \\
NGC1404        &0.86&0.82&0.41&/   &/   &/    \\
NGC1550        &0.72&0.59&0.49&/   &/   &/    \\
NGC3411        &0.54&1.36&1.04&/   &/   &/    \\                                                                                                 
NGC4261        &1.11&2.15&0.30&/   &/   &/    \\                                                                                                 
NGC4325        &0.60&1.17&0.62&/   &/   &/    \\                                                                                                 
NGC4374        &1.27&1.58&0.23&/   &/   &/    \\                                                                                                 
NGC4636        &0.78&0.68&0.33&/   &/   &/    \\                                                                                                 
NGC4649        &0.94&0.97&0.45&/   &/   &/    \\
NGC5813        &0.94&0.91&0.46&/   &/   &/    \\
NGC5846        &0.90&1.03&0.52&/   &/   &/    \\

\hline
\end{tabular}
\end{table*}

\bsp
\label{lastpage}

\end{document}